\documentstyle[bezier,11pt]{article}
\begin{document}
\def\Scri{
\unitlength=1.00mm
\thinlines
\begin{picture}(3.5,2.5)(3,3.8)
\put(4.9,5.12){\makebox(0,0)[cc]{$\cal J$}}
\bezier{20}(6.27,5.87)(3.93,4.60)(4.23,5.73)
\end{picture}}
{\hfill UMD PP 96-2}

{\hfill gr-qc/9507019}
\vskip3cm
\centerline{\Large\bf Aspects of Analyticity}
\vskip2cm
\centerline{\large \enspace Dieter R. Brill
\footnote{
Department of Physics, University of Maryland,
College Park, MD 20742, USA}}

\vskip4cm
\centerline{\large \bf CONTENTS}

\medskip
1 Introduction

\smallskip
2 Analytic Manifolds and Analytic Continuation of Metrics

\smallskip
3 Walker's Spacetimes and their Maximal Extension

\smallskip
4 Global Structure of de Sitter and Reissner-Nordstr\"om-de Sitter Cosmos

\hskip2cm 4.1 Special Cases

\hskip2cm 4.2 Collapsing Dust

\smallskip
5 Euclidean Metrics

\smallskip
6 Physical Interpretation of Euclidean Solutions, and a remark about the
Gravitational

\hskip4mmAction

\hskip2cm 6.1 Thermal Interpretation

\hskip2cm 6.2 Tunneling Interpretation

\smallskip
7 The Multi-Black-Hole Solutions

\hskip2cm 7.1 Merging Black Holes

\hskip2cm 7.2 Continuing Beyond the Horizons

\smallskip
8 Naked Singularities?

\smallskip
References

\newpage
\noindent
{\large\bf 1 Introduction}

\bigskip \noindent
In these lectures I discuss a number of diverse topics that happen to be
related to analyticity and analytic continuation. (This relation allows
for a concise title, but otherwise appears superficial --- I do not claim
to understand its the deeper basis, if any.)
I begin with some remarks about the uses and problems of analyticity
in spacetime physics. As an illustration of the possibilities
I discuss spherically symmetric spacetimes, whose analytic continuation
is completely understood. An example, the Reissner-Nordstr\"om-deSitter
space, is treated with particular attention to extreme cases.

Next I consider a different type of analytic continuation, namely to
``imaginary time", or Euclidean geometries. These may be interpreted
physically as thermal states or as tunneling states. As an example I
recall the thermal Schwarzschild and Reissner-Nordstr\"om solutions,
and the question of their entropy. This leads to a discussion of the
appropriate gravitational action, and the action associated with sharp
corners or ``joints" of the boundary. In the realm of tunneling states I
give several examples and their physical interpretation, with particular
attention to black hole pair creation and universe creation ``from
nothing."

Returning to Lorentzian analytic continuation I discuss the dynamical
multi-black-hole solutions of Kastor and Traschen.  This solution
describes a cosmology with several charged black holes in motion and
capable of collision. I explore the prospects of continuing the
solution beyond the region in which it was originally defined. The
spacetime so continued can then contain a naked singularity and provide a
counterexample to some versions of the cosmic censorship hypothesis.

\bigskip \bigskip \noindent
{\large\bf 2 Analytic Manifolds and Analytic Continuation of Metrics}

\bigskip \noindent
If we have a spacetime that can be covered by a single coordinate patch,
and the metric is an analytic function of those coordinates, then there
is usually no problem in extending the metric to the entire spacetime.
But interesting applications of Einstein's equations frequently occur
in manifolds of complicated topology, which cannot be covered by a
single coordinate system. In that case the region in which an exact solution
is first known is typically incomplete. The simplest way to complete it,
if possible, is by analytic extension. It is remarkable how little is
known in a systematic way about this frequently encountered problem.
Here we will not materially improve on this situation, but merely recall
some of what is known about analytic continuation, and discuss the most
common class of geometries for which a method exists.

It is tempting to expect that one can in general
continue analytic solutions (say of the sourceless Einstein equations)
beyond their initial coordinate neighborhood. However,
we will find that this is not always possible. That is, some solutions
that are analytic in one neighborhood and present no obvious obstruction
to continuation by way of diverging curvature invariants
may nonetheless be only of finite differentiability on the neighborhood's
boundary. There is of course nothing
unphysical about such behavior; it is what one expects, for example,
in the presence of a gravitational waves pulse of finite spatial extent.
The reason it is necessary to extend incomplete spacetimes is because
one purpose of spacetime is to describe the history of all inertial observers.
A (timelike) geodesically incomplete spacetime fails
to do this, so it behooves us to extend it as far as possible. (If even the
maximal extension is incomplete we can begin to ask questions about cosmic
censorship.) An analytic extension, if it exists, is special
because of its uniqueness and ``permanence".
The uniqueness is similar to that of the continuation of analytic functions:
If the metric is analytic in a neighborhood of an analytic spacetime, then
its analytic continuation is unique. If not only the metric
but also the manifold needs to be continued, then the continuation
is {\em not} necessarily unique. A simple example is a finite
part of (flat) Minkowski space, which can be continued either to the complete
Minkowski space, or to one of the several locally Minkowskian spaces, such
as the torus. More generally, any simply connected part of spacetime that
can be continued to a multiply connected one can also be continued to a
covering of the latter. (We will encounter such ambiguities with
cosmological black hole spacetimes in section 4). A somewhat more subtle
example is the
Taub-NUT space, which has two distinct and inequivalent analytic extensions
\cite{TN}. In these ambiguous cases one needs to decide on such properties
as the topology of the extended manifold along with the extension of the
metric. Once the whole (smooth) manifold is known, its analytic structure is
essentially unique.

How do we define an analytic manifold? To treat {\em real} extensions
the appropriate notion is real analyticity:
a function is analytic if it can be represented locally by a power series
expansion. On the other hand, we may want to consider complex extensions,
for example to ``imaginary time"; in that case we would use the usual
notion of analyticity in the complex plane. In either case a manifold is
analytic if the coordinate transformations between neighborhoods are
(real resp.\ complex) analytic functions. Tensors are analytic
if their components in analytic coordinates are analytic. Of course, in
non-analytic coordinates the components will in general not be analytic,
even if the tensor itself is.

Suppose we have an analytic differential equation, such as the sourceless
Einstein equations, on such a manifold, and a local analytic solution.
An analytic continuation of the solution will then also satisfy the
(continued) differential equation.\footnote{\small Analyticity
is not assured, nor is it necessarily
to be expected on physical grounds, if there are source terms present.
Well-known examples are stellar models that are non-analytic on the
stellar surface.} This is the ``permanence" property of analytic
continuations. It makes such continuations interesting because one gets
a ``new" solution ``for free," i.e. without having to solve the equation
again. But the continuation is not necessarily an easy matter
when the metric is given in some coordinates, and we desire to
extend across a boundary where both the metric coefficients and the coordinates
are non-analytic. There appears to be no systematic criterion for deciding
whether analytic continuation is possible, and one usually has to rely on
ingenuity to find suitable new coordinates in which the metric is
analytic in the relevant region.

It can also happen that neither the metric nor the coordinates are analytic
functions on the manifold, but the metric coefficients are analytic
functions of the coordinates; or they many be extendable to another (real)
range of the coordinates by an excursion in the complex plane. Examples
of this are found in the Schwarzschild metric at $r = 2M\/$ and $r = 0$
respectively. In either case we obtain solutions of the Einstein equations
in the new coordinate range, but it is a separate question whether and
how the geometry so described fits together with the original geometry,
and if so, whether the fit is analytic. Finding a proper overlap seems
to be the only way to assure the latter.\footnote{\small It is remarkable that
some solutions known in closed form (which is sometimes --- loosely --- called
``analytic") can be extended with a high degree of differentiability
(e.g. $C^{122}$ as in \cite{Cru}, and in certain of the spaces discussed
in section 7) but not analytically.} For a class
of metrics, which I discuss in section 3, one knows how to fit together
the pieces across horizons like $r = 2M$ for the Schwarzschild geometry.

One can, of course, give criteria that establish {\it non\/}-analyticity,
for example the divergence of invariants formed from the Riemann tensor
and/or its derivatives. This happens, for example, at $r = 0$ in the
Schwarzschild metric, so no real analytic extension is possible there.
But for indefinite metrics not all divergences of the Riemann tensor
can be found in this way. Namely, the Riemann tensor may have
a ``null" infinity, as in $R_{\mu \nu \alpha \beta} =
l_{[\mu}m_{\nu]}l_{[\alpha}m_{\beta]}$,
with $l^\mu l_\mu = 0$, $l^\mu m_\mu = 0$ and divergence in $l,\/m$. In
that case the divergence can be found by evaluating the Riemann tensor
components in an orthonormal frame. (To avoid spurious
infinities due to the frame becoming null the orthonormal frame should be
parallely propagated \cite{ES}).

If one admits an excursion to complex coordinate values one may find
other real metrics analytically related to the original one. Because there
may be other ``real sections" of the complex metric, such extensions
may not be unique.\footnote{\small To restore uniqueness it has been
suggested \cite{PSH} that the slightly complex path should be a
geodesic. It remains to be seen whether one does not still lose
physical significance in this unorthodox continuation.}
Also, they may have nothing directly to do with the original geometry.
In fact, the ``extension" may have a different signature than the
original metric. This is a favorite way to generate Euclidean solutions
such as the ones discussed in section 5. An example  of
a Lorentzian, complex analytic relation of the
Schwarzschild metric (which can however hardly be said to be physically
related) is
$$ds^2 = \left(1-{2M\over r}\right)dt^2 +
{dr^2\over{\left(1-{2M\over r}\right)}} - r^2 d\theta^2 +
r^2 \cosh^2\!\theta\, d\phi^2.$$
(This is obtained from the usual Schwarzschild geometry by
$t \rightarrow it, \, {\pi\over 2} -\theta \rightarrow i\theta$.)
Another example is the continuation of the de Sitter space metric,
$$ds^2 = - dt^2 + e^{2Ht}(dx^2 + dy^2 + dz^2)$$
across $t = 0$, which effectively
changes the cosmological expansion parameter $H$ into its negative.

Because ``new" information can propagate along null surfaces, such surfaces
are a natural analyticity boundary. On the other hand,
analyticity of a region would be expected to extend to the domain of dependence
of that region. Even when the geometry itself can be extended, the coordinates
in which a metric is originally found tend to be analytic only in such domains
of dependence.  One of the few, somewhat general approaches to extend such
metrics proceeds by introducing null coordinates in which the boundary is
one of the coordinate surfaces. Such is the case in the following class of
metrics.

\bigskip \bigskip \noindent
{\large\bf 3 Walker's Spacetimes and their Maximal Extension}

\bigskip \noindent
The extensions of spherically symmetric ``static" metrics of the form
\begin{equation}
ds^2 = -F dt^2 + {dr^2 \over F} + r^2 d\Omega^2
\end{equation}
have been completely discussed by Walker \cite{WA}.
Here $F = F(r)$ is the norm of the Killing vector $\partial/\partial t$;
it is not required to be timelike (hence the quotes
around ``static") because we allow $F$ to be positive or negative.
$F$ may be an analytic function of $r$, satisfying the Einstein equation,
and range over positive and negative values,
but the metric is clearly non-analytic at the zeros of $F$; the
problem is to find the continuation across these zeros.
(Infinities of $F$ imply infinities of the Riemann tensor, so no analytic
continuation is possible there.) Because the angular part is regular
for $r > 0$, it suffices to confine attention to the two-dimensional
$r,\,t$ part of the metric.

In this two-dimensional space one can easily introduce null coordinates
by ``factoring" the metric as a product of two integrable null
differential forms,
\begin{equation}
 du = dt + {dr \over F}\/, \qquad dv = dt - {dr \over F}.
\end{equation}
The metric then takes the double null form, $ds^2 = -F(u\!-\!v)\,dudv$,
but this is still singular at $F = 0$. If instead, following
Finkelstein's trick, we use $r$ and only {\em one} null coordinate, say $u$,
the metric assumes the nonsingular form
$$ds^2= -F(r) du^2 + 2 du dr,$$
which is analytic wherever $F$ is analytic as a function.\footnote{\small
If $F$ is only
smooth or $C^n$, this provides a smooth or $C^n$ extension across $F=0$.
In those cases the extension is of course not unique.}  This metric, then,
provides the overlap
necessary to connect two regions with opposite signs of $F$.
How two such regions fit together can be shown
by a conformal diagram, as in Fig.\ 1.
Here we have assumed that the region $r \rightarrow \infty$ has the usual
asymptotically flat structure, and that there is another zero of $F$
at finite $r$ below $r=a$ (corresponding to the ``roof'' of the figure).
If these structures are different, the blocks may be shaped differently,
but the region around the zero, $r=a$, will look the same.

\bigskip\bigskip

\unitlength=.75mm
\thicklines
\begin{picture}(120.00,100.00)(0,20)
\multiput(0,0)(-30,30){3}{\put(90.00,30.00){\line(1,1){30.00}}}  
\multiput(0,0)(-30,-30){2}{\put(120.00,60.00){\line(-1,1){60.00}}} 
\multiput(0,0)(-30,30){2}{
\bezier{288}(60.00,60.00)(90.00,40.00)(120.00,60.00)} 
\put(110.00,50.00){\line(-1,1){60.00}}  
\bezier{288}(90.00,90.00)(70.00,60.00)(90.00,30.00)  
\bezier{288}(60.00,60.00)(90.00,80.00)(120.00,60.00)  
\thinlines
\put(112.33,74.67){\vector(-4,-3){8.33}}
\put(112.33,74.67){\vector(0,-1){19}}
\put(122.0,76.00){\makebox(0,0)[cc]{$t=$ const}}
\put(90.33,22){\makebox(0,0)[cc]{$t = -\infty$}}
\put(41.00,70.33){\vector(2,1){18.70}}
\put(45.00,63.33){\vector(4,1){23.70}}
\put(45.00,63.33){\vector(1,1){5.60}}
\put(49.67,55.67){\vector(1,0){30.00}}
\put(27,70){\makebox(0,0)[cc]{$r=$ const $<a$}}
\put(30.00,62){\makebox(0,0)[cc]{$r=a$, $F=0$}}
\put(35.5,55.50){\makebox(0,0)[cc]{$r=$ const $>a$}}
\put(93.00,99.33){\vector(-3,-1){24}}
\put(103,101.00){\makebox(0,0)[cc]{$u =$ const}}
\put(103.67,84.67){\vector(0,-1){7.30}}
\put(103.67,84.67){\vector(-1,0){18.67}}
\put(113.5,85.50){\makebox(0,0)[cc]{$t=+\infty$}}
\put(90.00,60.00){\makebox(0,0)[cc]{$F>0$}}
\put(60.00,90.00){\makebox(0,0)[cc]{$F<0$}}
\put(84.67,24.00){\vector(-1,4){3.67}}
\put(96.00,24.00){\vector(1,4){3.67}}
\end{picture}

\bigskip

{\noindent \small
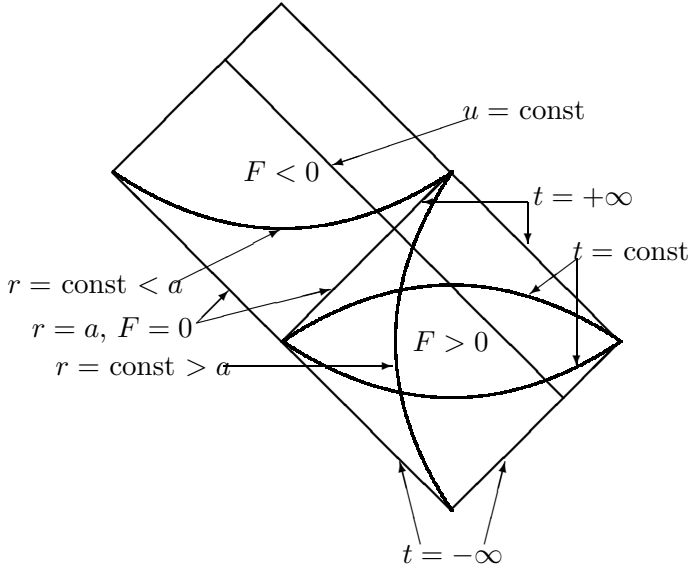
{\bf Fig.\ 1.} Conformal diagram of region of Walker metric
surrounding $r=a$, the largest zero of $F$. The $r,t$
coordinates are degenerate on the boundaries of the diamond-shaped
regions. For example, along the lower left boundary both $t$ and $r$
are constant (namely $r=a$).}

\bigskip\bigskip
The spacetime as shown is still not complete. For example, the middle left
boundary labeled $r=a$ is at a finite distance from typical points
inside the region, and so is the ``roof." By using
Finkelstein coordinates $r$ and $v$ we obtain a system that overlaps the
middle left boundary, and by repeating this procedure around the next
smaller zero below $r=a$ we can extend beyond the ``roof." \, Thus in passing
through each zero of $F$ we add several new diamond-shaped regions
to the conformal diagram, according as we cross the boundary along $u =$ const
or $v =$ const. All the regions so generated at the zero of $F$ where $r=a$
are shown in Fig.\ 2.

\unitlength=.60mm
\begin{picture}(130.00,141.00)(-15,20)
\thicklines
\put(70.00,20.00){\line(1,1){60.00}}
\put(130.00,80.00){\line(-1,1){60.00}}
\put(70.00,140.00){\line(-1,-1){60.00}}
\put(10.00,80.00){\line(1,-1){60.00}}
\put(40.00,50.00){\line(1,1){60.00}}
\put(100.00,50.00){\line(-1,1){60.00}}
\thinlines
\put(70.00,81.00){\makebox(0,0)[cb]{P}}
\put(101.00,111.00){\makebox(0,0)[cb]{Q}}
\put(70.00,141.00){\makebox(0,0)[cb]{R}}
\put(116.00,96.00){\makebox(0,0)[lb]{$r=\infty,\,t=+\infty$}}
\put(118.00,66.00){\makebox(0,0)[lt]{$r=\infty,\,t=-\infty$}}
\put(82.00,30.00){\makebox(0,0)[lt]{$r=b,\,t=-\infty$}}
\put(94.00,40.00){\makebox(0,0)[lt]{$r=a,\,t=-\infty$}}
\put(94.00,40.00){\vector(-1,3){7.33}}
\put(112.00,106.00){\makebox(0,0)[lc]{$r=a,\,t=+\infty$}}
\put(112.00,106.00){\vector(-3,-1){22.00}}
\put(90.00,124.00){\makebox(0,0)[lb]{$r=b,\,t=+\infty$}}
\put(50.00,122.00){\makebox(0,0)[rb]{$r=b,\,t=-\infty$}}
\put(24.00,96.00){\makebox(0,0)[rb]{$r=\infty,\,t=-\infty$}}
\put(24.00,64.00){\makebox(0,0)[rt]{$r=\infty,\,t=\infty$}}
\put(52.00,34.00){\makebox(0,0)[rt]{$r=b,\,t=+\infty$}}
\put(100.00,80.00){\makebox(0,0)[cc]{$F>0$}}
\put(70.00,50.00){\makebox(0,0)[cc]{$F<0$}}
\put(40.00,80.00){\makebox(0,0)[cc]{$F>0$}}
\put(70.00,110.00){\makebox(0,0)[cc]{$F<0$}}
\end{picture}

\bigskip
{\noindent \small
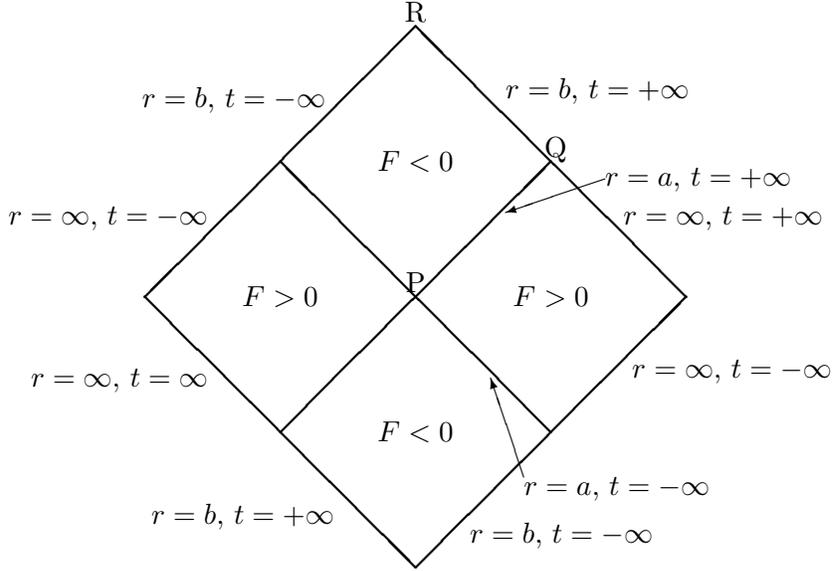
{\bf Fig.\ 2.} Blocks that fit together at their $r=a$ boundary. The
next lower zero of $F$ occurs at $r=b$.}

\bigskip\bigskip

The four blocks in Fig.\ 2 are connected across all the diagonal lines,
because the overlapping coordinates constructed so far reach across all of
these lines. But these coordinates are not good at the intersection points
P, Q, R, \ldots, so we do not yet know whether the regions fit together
at a point like P, where four regions meet. In fact, the intersection
points come in two types: those like P and R are characterized by the
vanishing of the $\partial/\partial t$ Killing vector, whereas at points like
Q different values of $r$ are ``trying to come together." \, It is therefore
not surprising that no analytic continuation is possible
or necessary across points
of type Q; they lie at an infinite distance ($t \rightarrow \infty$)
along Killing orbits and are not part of the manifold.

At points of type P the blocks may fit together smoothly
or analytically, depending on the form of $F$. We cannot analyze point P
using the coordinates $u$ or $v$, because we have $u = -\infty, \,
v = \infty$ there. But if we introduce exponentials of these coordinates,
$$U = e^{cu}, \qquad V = e^{-cv}$$
where $c$ is a adjustable constant, then (for $c>0$) $U = 0 = V$ at P
and the metric takes the form
$$ds^2 = {F\over c^2}\,e^{\left(-2c \int {dr\over F}\right)}\, dU dV.$$
The trick is now to choose $c$ so that the conformal factor in this
expression is finite. It is not difficult to verify that this can be
done when $F$ has a simple zero (for example, if $F = r - r_0$ one finds
$c={1\over 2}$). For functions $F$ of the type
\begin{equation}
F(r) = {\prod_i (r-a_i)\over K(r)}\, ,
\end{equation}
with $K(r)$ a polynomial with zeros differing from the $a_i$,
Walker \cite{WA} shows that $c$ can be
chosen so that the metric becomes regular at P --- the four blocks fit
together smoothly or analytically.

If two roots coincide, the picture looks different.
There are no points of type P because the double-root horizon is
at an infinite spatial distance; a spacelike section does not have
the ``wormhole'' shape, but is an infinite funnel or ``cornucopion.''
Figure 3 shows how the blocks fit together in that case \cite{KL}.
All lines and curves that are shown correspond to $r =$ const.
This diagram looks like what one would obtain by continuing Fig.\ 1
towards the upper left by the usual rules to another block with $F>0$,
and then eliminating the $F<0$ block and moving the two $F>0$ blocks
together. Thus the coincidence limit of two roots of $F$ does not
appear continuous in the conformal picture.

\medskip

\unitlength=.50mm
\thicklines
\begin{picture}(120.00,120.00)(-30,20)
\multiput(0,0)(-60,60){2}{\put(90.00,30.00){\line(1,1){30.00}}}  
\put(120.00,60.00){\line(-1,1){60.00}} 
\bezier{288}(60.00,120.00)(80.00,90.00)(60.00,60.00)  
\bezier{288}(90.00,90.00)(70.00,60.00)(90.00,30.00)  
\thinlines
\multiput(0,0)(.5,.5){2}{
\put(89.50,29.50){\line(-1,1){60.00}}}
\multiput(0,0)(-.5,.5){2}{\put(60.50,59.50){\line(1,1){30.00}}}
\put(95.00,100.00){\vector(-1,0){25.00}}
\put(96.00,100.00){\makebox(0,0)[lc]{$r =$ const $<a$}}
\put(120.00,75.00){\vector(-1,0){37.00}}
\put(122.00,75.00){\makebox(0,0)[lc]{$r=$ const $>a$}}
\end{picture}

{\noindent \small
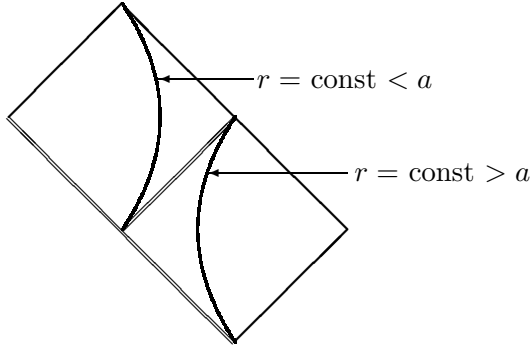
{\bf Fig.\ 3.} Conformal diagram of region of Walker metric surrounding
a double root of $F$, denoted by double lines.}

\bigskip\bigskip

This happens, for example,
for the Reissner-Nordstr\"om geometry. For this case one has
$$F = 1 - {2M\over r} + {Q^2\over r^2} = {(r-r_+)(r-r_-)\over r^2}.$$
The coincidence limit $r_+ \rightarrow r_-$
corresponds to an ``extremally'' charged black hole, $Q^2 \rightarrow
M^2$. As long as the roots are
distinct the two $F>0$ blocks have a finite size $F<0$ block between
them, but at the extremal limit the situation of Fig. 3 applies
(except that the upper left block ends at $r = 0$ rather than
continuing to another zero of $F$).

One can also take a different limit, which leads to another solution
that has nothing directly to do with the analytic continuation of the
first block.
Instead of eliminating the $F<0$ block, one can keep its physical size
constant by rescaling the metric: define $a$ and $\phi$ by
$$2a = r_+ - r_-, \qquad a \cos (\phi/a) = r - {1\over 2}(r_+ + r_-).$$
For small $a$ the region between the two roots has the metric
$$ds^2 = (a/M)^2 \sin^2\phi\, dt^2 - (aM)^2 d\phi^2.$$
The metric $ds^2/a^2$ is related to the
Bertotti-Robinson universe. (More detail on the relation between the
extremal Reissner-Nordstr\"om and the Bertotti-Robinson geometries is
found in \cite{CABR}.)

Beyond the coincidence limit it can happen that a pair of (real) roots
disappear. This is also not a continuous change in the conformal
diagram --- the two regions of Fig.\ 3 merge into one.

\bigskip \bigskip

{\large\bf\noindent
4 Global Structure of de Sitter and Reissner-Nordstr\"om-de Sitter Cosmos}

\bigskip \noindent
The Reissner-Nordstr\"om-de Sitter (RNdS) metric for
a black hole of mass $M$ and charge $Q$ in a
universe with cosmological constant $\Lambda$ is of type (1), with $F$ of the
form
\begin{equation}
F(r) = {\left(- {\textstyle {1 \over 3}}\Lambda  r^4 + r^2 -
2Mr + Q^2\right)\over r^2}\, ,
\end{equation}
which is of the type (3).  Thus we know that
the maximal analytic extension is given by the Walker construction, and
once we know how the blocks fit together we can do all calculations in
the original $r$, $t$ coordinates.

{}From the above block-gluing rules it is clear that the conformal diagrams
for the Reissner-Nordstr\"om-de Sitter metrics depend only on the
number of zeros of the function $F$ in (4). We will denote the zeros of the
numerator, in decreasing order, by $a_1$ \ldots $a_4$. Only three of the roots
are positive (for positive $M$ and $\Lambda$). The simplest example is
de Sitter space itself, $M = 0$, $Q = 0$, so $a_1 = \sqrt{3/\Lambda}$,
$a_2=a_3=0$. The blocks look different in this case, because $r=0$ is
a regular origin. Also, $r=\infty$ is infinite distance in time (since
$F < 0$ for $r > a_1$), so we can identify it with timelike and null
infinity, \Scri. Figure 4a shows an embedding of an $r,\,t$ subspace of
this geometry in flat 3-dimensional Minkowski space, and Fig.\ 4b is the
corresponding conformal diagram. Note that the conformal diagram corresponds
to only half of the embedded surface, because the latter shows both ``sides''
of the origin ($\phi=0$ and $\phi=\pi$, for example).

\bigskip

\unitlength=0.80mm
\begin{picture}(125.00,73.00)(-5,0)
\thinlines
\put(10.00,10.00){\line(1,1){40.00}}
\put(10.00,50.00){\line(1,-1){40.00}}
\bezier{100}(42.00,32.00)(30.00,28.00)(18.00,32.00)
\thicklines
\bezier{220}(52.00,11.00)(33.00,30.00)(52.00,51.00)
\bezier{228}(8.00,51.00)(28.00,30.00)(8.00,11.00)
\bezier{196}(8.00,11.00)(30.00,0.00)(52.00,11.00)
\thinlines
\put(36.00,47.00){\vector(1,4){2.67}}
\put(39.00,60.00){\makebox(0,0)[cb]{$r \rightarrow \infty\,\,(\Scri^+)$}}
\put(52.00,30.00){\vector(-2,-1){9.00}}
\put(53.00,31.00){\makebox(0,0)[lb]{$r=0$}}
\put(51.00,28.00){\makebox(0,0)[lc]{(antipodal}}
\put(52.00,26.00){\makebox(0,0)[lt]{observer)}}
\put(7.00,43.00){\vector(1,0){10.00}}
\put(6.00,43.00){\makebox(0,0)[rc]{$r=a_1$}}
\put(5.00,30.75){\makebox(0,0)[rc]{$t=0$}}
\put(6.00,22.00){\vector(1,0){10.00}}
\put(5.00,22.00){\makebox(0,0)[rc]{$r=0$}}
\put(8.00,18.00){\makebox(0,0)[rc]{(observer)}}
\put(6.00,30.50){\vector(1,0){15.00}}
\put(85.00,10.00){\line(1,1){40.00}}
\put(85.00,10.00){\framebox(40.00,40.00)[cc]{}}
\put(125.00,10.00){\line(-1,1){40.00}}
\put(105.00,51.00){\makebox(0,0)[cb]{$r=\infty \enspace \Scri^+$}}
\put(96.00,28.90){\makebox(0,0)[cb]{$t=0$}}
\put(84.00,32.00){\makebox(0,0)[rc]{$r=0$}}
\put(126.00,32.00){\makebox(0,0)[lc]{$r=0$}}
\put(95.00,40.00){\makebox(0,0)[cc]{$r=a_1$}}
\put(115.00,40.00){\makebox(0,0)[cc]{$r=a_1$}}
\put(105.00,10.00){\makebox(0,0)[ct]{$r=\infty \enspace \Scri^-$}}
\multiput(0,0)(5,0){8}{\put(86.00,30.00){\line(1,0){3.00}}}
\put(30.00,0.00){\makebox(0,0)[cb]{a}}
\put(105.00,0.00){\makebox(0,0)[cb]{b}}
\thicklines
\bezier{32}(9.83,50.17)(5.83,51.17)(9.83,52.17)
\bezier{168}(9.83,50.17)(30.00,43.50)(50.00,50.00)
\bezier{164}(9.83,52.17)(30.17,56.50)(50.17,52.17)
\bezier{28}(50.00,50.00)(53.67,51.33)(50.17,52.17)
\end{picture}

\bigskip
{\noindent \small
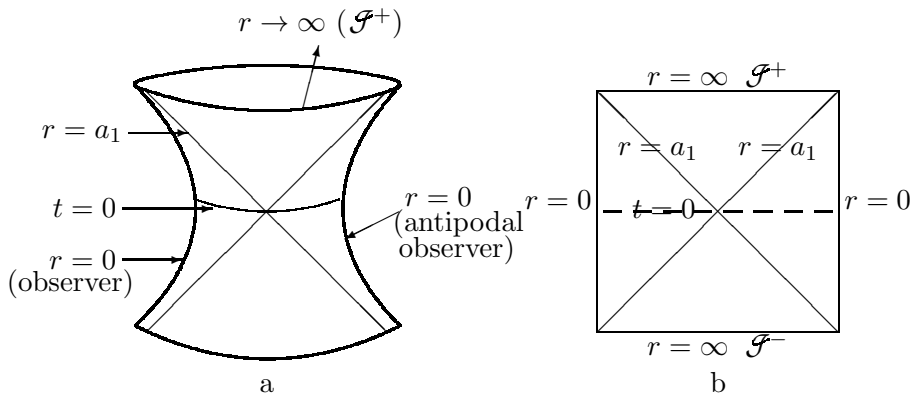
{\bf Fig.\ 4a.} Embedding of 2-D de Sitter space in 3-D
Minkowski space.

\noindent {\bf Fig.\ 4b.} The corresponding conformal diagram.}

\bigskip\bigskip

De Sitter space is often described in coordinates different from the
$r,\,t$ used above, in which the space sections are conformally flat.
These new coordinates $r',\, t'$ are called {\em isotropic} or
{\em cosmological} coordinates \cite{BH},
$$r = r'e^{Ht'} \qquad t = t' - {1\over 2H}\ln (1- H^2r^2).$$
Here $H = \sqrt{\Lambda/3} = 1/a_1$ is the ``Hubble constant" or
cosmological expansion parameter. The metric then becomes
\begin{equation}
ds^2 = -dt'^2 + e^{2Ht'}\left(dr^2 + r^2 d\Omega^2\right).
\end{equation}
These coordinates cover more of de Sitter space than one patch of the
$r,\,t$ coordinates, but there still is a horizon at $t' = -\infty$.
Another block of $r',\,t'$ coordinates  but with opposite sign of
$H$ can be analytically connected to the original one.  We call the
coordinates ``expanding'' in the block with $H > 0$, and ``contracting''
in the other one. Figure 5 shows some of the spacelike surfaces $t'=$ const.

\unitlength=.70mm
\begin{picture}(105.00,85.00)(-10,5)
\thinlines
\put(45.00,10.00){\framebox(60.00,60.00)[cc]{}}
\bezier{256}(45.00,10.00)(75.00,25.00)(105.00,25.00)
\bezier{300}(45.00,10.00)(90.00,50.00)(105.00,50.00)
\thicklines
\put(45.00,10.00){\line(0,1){60.00}}
\put(45.00,70.00){\line(1,0){60.00}}
\put(45.00,10.00){\line(1,1){60.00}}
\bezier{300}(45.00,30.00)(60.00,30.00)(105.00,70.00)
\bezier{256}(105.00,70.00)(75.00,55.00)(45.00,55.00)
\thinlines
\put(65.00,50.00){\makebox(0,0)[cc]{$H>0$}}
\put(85.00,30.00){\makebox(0,0)[cc]{$H<0$}}
\put(75.00,71.00){\makebox(0,0)[cb]{$t'=\infty\enspace\Scri^+$}}
\put(75.00,10.00){\makebox(0,0)[ct]{$t'=\infty\enspace\Scri^-$}}
\put(40.00,25.00){\vector(4,1){26.00}}
\put(38.00,25.00){\makebox(0,0)[rc]{$t'= -\infty$}}
\put(40.00,43.00){\vector(4,-1){23.00}}
\put(38.00,43.00){\makebox(0,0)[rc]{$t'=$ const}}
\end{picture}

\bigskip
{\noindent \small
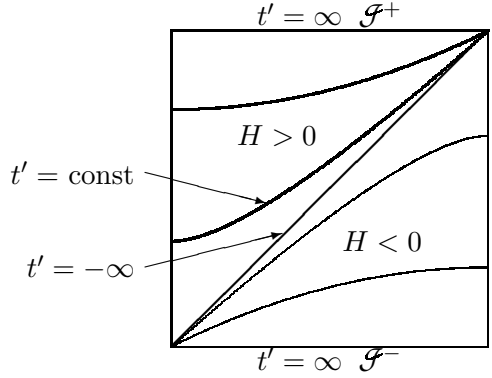
{\bf Fig.\ 5.} Cosmological coordinates in de Sitter space. The part drawn
in thick lines is the expanding region.}

\bigskip\bigskip

That the de Sitter universe can appear as either static (4), or expanding
or contracting (5), is an accident due to the high degree of symmetry
of this model. In fact, each ``observer" (timelike geodesic)
has a static frame centered around him/her. When there is a ``single''
black hole present, a static frame still exists (but only the one that
is centered about the black hole). We will therefore treat the analytic
extension of this case by Walker's method; but we also want to show how the
expanding and contracting, cosmological frames fit together
(because only their analog exists in the spacetimes \cite{KT}
with several black holes.)

We rewrite the RNdS metric (4) using capital letters to denote the
static frame (in order to distinguish it from the cosmological coordinates,
which will be in lower case):
$$ds^2 = -F\,dT^2 + {dR^2\over F} + R^2\,d\Omega^2$$
$$F = 1-{2m\over R}+{Q^2\over R^2}-{1\over 3}\Lambda R^2 \qquad
A_T = -{Q\over R}\,.$$
By means of a somewhat involved coordinate transformation
\cite{BH} one finds the form of the metric in cosmological coordinates,
\begin{equation}
ds^2 = -V^{-2} dt^2 + U^2 e^{2Ht} (dr^2 + r^2 d\Omega^2)\,,
\end{equation}
where
$$U = 1+{M\over\rho}+{M^2-Q^2\over 4\rho^2} \qquad
V={U\over 1-{M^2-Q^2\over 4\rho^2}} \qquad \rho=e^{Ht} r\,.$$
By means of the simple coordinate change
$$\tau \equiv H^{-1}e^{Ht}$$
we can extend the region covered by these coordinates, by
allowing $\tau$ to be negative as well as positive. The metric
and EM potential then become
\begin{equation}
ds^2=-{d\tau^2\over U^2}+U^2(dr^2+r^2d\Omega^2) \qquad
U = H\tau+{M\over r} \qquad A_\tau={1\over U}\,.
\end{equation}
Figure 6 shows the analytic extension constructed according to the
prescription of Sect.\ 3 for the generic case, when there are three
roots of $F$.

\unitlength=.5mm
\begin{picture}(60.00,145.00)(-30,-4)
\thinlines
\multiput(0,0)(0,40){2}{
    \put(20.00,40.00){\line(1,0){40.00}}}
\put(100.00,40.00){\line(1,0){40.00}}
\multiput(0,0)(40,0){2}{
    \multiput(0,0)(0,-3){13}{
        \put(60.00,120.00){\makebox(0,0)[ct]{*}}}}
\multiput(0,0)(0,-3){13}{
    \put(60.00,40.00){\makebox(0,0)[ct]{*}}}
\multiput(0,0)(0,-3){13}{
    \put(100.00,40.00){\makebox(0,0)[ct]{\bf*}}}
\multiput(0,0)(80,0){2}{
    \put(40.00,39.00){\makebox(0,0)[ct]{$\Scri^-$}}
    \put(40.00,81.00){\makebox(0,0)[cb]{$\Scri^+$}}}
\multiput(-40,0)(80,0){2}{        
   \multiput(0,0)(2,-2){20}{
        \put(60.00,80.00){\makebox(0,0)[cb]{.}}}}
\multiput(0,0)(0,40){2}{          
   \multiput(0,0)(2,-2){20}{
        \put(60.00,80.00){\makebox(0,0)[cb]{.}}}}
\multiput(0,0)(2,2){40}{          
        \put(20.00,40.00){\makebox(0,0)[cb]{.}}}
\multiput(0,0)(2,2){20}{          
        \put(60.00,0.00){\makebox(0,0)[cb]{.}}}
\thicklines
\put(100.00,80.00){\line(1,0){40.00}}
\put(60.00,40.00){\line(1,1){40.00}}
\put(100.00,40.00){\line(1,1){40.00}}
\put(60.00,40.00){\line(1,-1){40.00}}
\bezier{180}(60.00,40.00)(80.00,50.00)(100.00,40.00)
\bezier{180}(60.00,40.00)(80.00,30.00)(100.00,40.00)
\bezier{92}(90.00,48.00)(80.00,42.00)(70.00,48.00)
\bezier{244}(90.00,48.00)(130.00,80.00)(140.00,80.00)
\bezier{120}(70.00,30.00)(80.00,40.00)(92.00,30.00)
\bezier{40}(92.00,30.00)(96.00,26.00)(100.00,26.00)
\thinlines
\bezier{244}(70.00,48.00)(30.00,80.00)(20.00,80.00)
\bezier{48}(60.00,26.00)(66.00,26.00)(70.00,30.00)

\end{picture}

\smallskip
{\noindent \small
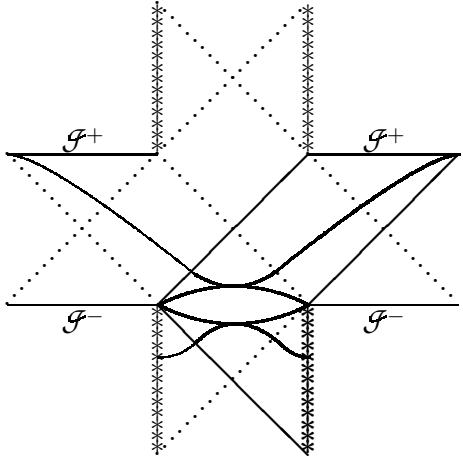
{\bf Fig.\ 6.} Conformal diagram of the RNdS geometry. The diagonal
(mostly dotted) lines are the horizons corresponding to the
three roots of $F$ that separate the different
``static'' blocks. Those crossing between $\Scri^-$ and $\Scri^+$ are
the cosmological horizons at $R=R_c=a_1$; those crossing at the center of the
figure are the outer black hole horizons at $R=R_o=a_2$; and those crossing at
the top and at the bottom are the inner horizons at $R=R_i=a_3$.
The multiply-crossed
vertical lines are the singularities at $R=0$. The thin curves
describe surfaces of constant cosmological time $\tau$, the upper
one having $\tau>0$, and the lower one $\tau<0$. The cosmological
$r$-coordinate is single-valued only to the right (or only to the left)
of the point of contact of these curves with the lens-shaped region.
This point of contact is a ``wormhole throat'' of the spacelike
geometry. Surfaces of different $\tau$-value touch the lens-shaped
region at different points. The boundaries of the lens-shaped region
are given by $R={\rm const}=M\pm\sqrt{M^2-Q^2}$.
The two region covered by the right-hand
parts of these spacelike surfaces, in which the
cosmological coordinates are therefore single-valued, is shown by the
thick outlines. Note that the part near $\Scri^+$ is similar to
that shown in Fig.\ 5. The
lens-shaped region in between is not covered by these coordinates. These
regions as drawn are appropriate for expanding coordinates. A region
covered by contracting coordinates is obtained, for example, by
reflecting the regions in thick outline about the horizontal symmetry
axis.}

\bigskip\bigskip
The figure can be repeated indefinitely in the horizontal and the
vertical direction, yielding a spacetime that is spacelike and
timelike periodic. The spacelike surface $T=0$ that cuts the
figure horizontally in its center then has the geometry akin to
a string of beads: its spatial geometry is given by
$$ds^2 = F^{-1}dR^2+R^2dS^2,$$
which can be embedded in four-dimensional flat space,
$dZ^2+dR^2+R^2dS^2$ by
$$Z=\int\sqrt{F^{-1}-1}\,dR.$$
The surface has maximal radius $R_c$ at the cosmological horizon,
and minimal radius $R_o$ at the throats of the holes.
The maxima correspond to the large
regions of the universe (the ``background de Sitter space''),
and the minima are throats of wormholes that connect one de Sitter
region with the next. Thus each de Sitter region contains two
wormhole mouths, placed in antipodal regions of each large universe.
(This is the reason for the quotes above when calling this a
``single'' black hole in a de Sitter universe.) One of these de Sitter
regions is shown in Figure 7.
Isometric copies of this surface can be smoothly joined at the throats,
producing a periodic $S^2\times R^1$ spatial topology,
in which the interiors of the black holes lead to further de Sitter regions.

\unitlength 1.00mm
\begin{picture}(81.33,70.00)(0,15)
\thicklines
\bezier{64}(35.86,64.14)(41.72,70.00)(50.00,70.00)
\bezier{68}(35.86,64.14)(26.67,53.67)(22.33,54.00)
\bezier{44}(22.33,46.00)(18.67,50.00)(22.33,54.00)
\bezier{64}(64.14,64.14)(58.28,70.00)(50.00,70.00)
\bezier{68}(64.14,64.14)(73.33,53.67)(77.67,54.00)
\bezier{44}(77.67,46.00)(81.33,50.00)(77.67,54.00)
\bezier{40}(77.33,53.67)(74.00,50.00)(77.33,46.00)
\bezier{64}(35.86,35.86)(41.72,30.00)(50.00,30.00)
\bezier{68}(35.86,35.86)(26.67,46.33)(22.33,46.00)
\bezier{64}(64.14,35.86)(58.28,30.00)(50.00,30.00)
\bezier{68}(64.14,35.86)(73.33,46.33)(77.67,46.00)
\thinlines
\bezier{64}(34.86,57.14)(40.72,60.10)(49.00,60.10)
\bezier{68}(34.86,57.14)(25.67,51.85)(21.33,52.02)
\bezier{64}(63.14,57.14)(57.28,60.10)(49.00,60.10)
\bezier{68}(63.14,57.14)(72.33,51.85)(76.67,52.02)
\bezier{64}(33.86,42.50)(39.72,39.39)(48.00,39.39)
\bezier{68}(33.86,42.50)(24.67,48.05)(21.00,47.88)
\bezier{64}(62.14,42.50)(56.28,39.39)(48.00,39.39)
\bezier{68}(62.14,42.50)(71.33,48.05)(75.67,47.88)
\bezier{64}(33.86,48.44)(39.72,47.37)(48.00,47.37)
\bezier{68}(33.86,48.44)(24.67,50.33)(20.33,50.27)
\bezier{64}(62.14,48.44)(56.28,47.37)(48.00,47.37)
\bezier{68}(62.14,48.44)(71.33,50.33)(75.67,50.27)
\bezier{188}(50.00,69.99)(37.67,47.60)(50.00,30.00)
\put(21.83,45.20){\vector(0,1){0.2}}
\put(21.83,35.20){\line(0,1){10.00}}
\put(49.50,28.80){\vector(0,1){0.2}}
\put(49.50,22.40){\line(0,1){6.40}}
\put(77.33,44.80){\vector(0,1){0.2}}
\put(77.33,35.20){\line(0,1){9.60}}
\put(21.83,31.60){\makebox(0,0)[cc]{$R=R_o$}}
\put(77.33,32.00){\makebox(0,0)[cc]{$R=R_o$}}
\put(49.50,20.00){\makebox(0,0)[cc]{$R=R_c$}}
\end{picture}

{\noindent \small
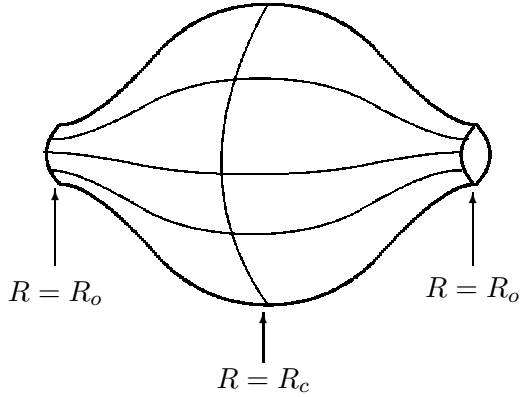
{\bf Fig.\ 7.} The geometry of the $T = 0$ slice of Fig.\ 6, embedded
in flat space with a polar angle suppressed. Typical electric field
lines are shown.}

\bigskip\bigskip
Alternatively and more compactly we can imagine the left and right
halves of the figure identified, so that the horizontal spacelike
surface of the figure is a closed circle, and the 3-D spacelike
topology is $S^1 \times S^2$. In this case the electric flux of
the charge $Q$ also describes closed circles: it emerges from one wormhole
mouth, spreads out to the maximum universe size, reconverges on the
other mouth, and flows through the wormhole back to the first mouth.
Seen from the large universe, the first mouth appears positively
charged, and the antipodal one, negatively charged --- an example
of Wheeler's \cite{JAW} ``charge without charge''!

\bigskip \bigskip \noindent
{\bf 4.1 Special cases}

\bigskip \noindent
The picture of Fig.\ 6 changes, for example as in Fig.\ 3, when roots
of $F$ coincide or cease to be real (see \cite{BH} and \cite{K?}).
It is easily checked that the quartic $R^2F(R)$ has exactly one negative root,
and generically either one or three positive roots,
with special cases of double or triple roots.
The possibilities for the positive roots are as follows.

\noindent(i) The generic black-hole case: three simple roots $R_c>R_o>R_i$,
as in Fig.\ 6. There are three types of Killing horizon:
cosmological horizons at $R=R_c$ and inner and outer black-hole horizons
at $R=R_i$ and $R=R_o$ respectively.

\noindent(ii) The generic naked-singularity case: one simple root,
giving only cosmological horizons (as in Fig.\ 9).

\noindent(iii) The extreme (or ``cold'') black-hole case:
a double root $R_d$ and a simple root $R_c>R_d$,
corresponding to a degenerate black-hole horizon at $R=R_d$
and a cosmological horizon at $R=R_c$.

\noindent(iv) The extreme (or marginal) naked-singularity case:
a double root $R_d$ and a simple root $R_i<R_d$,
corresponding to a degenerate cosmological horizon at $R=R_d$
and an inner horizon at $R=R_i$.

\noindent(v) The ``ultra-extreme'' case:
triple root, with one doubly degenerate horizon.

\def \L{\Lambda}
To make explicit which case occurs for each $(M,Q,\L)$, we note
that the double roots occur if and only if
$M=M_\pm(Q,\L)$, where
$$M_\pm=P_\pm(1-{2\over 3}\L P_\pm^2),\qquad
P_\pm^2={1\over{2\L}}\left(1\pm\sqrt{1-4\L Q^2}\right),$$
with the double root being at $R=P_\pm$ \cite{RO}.
The ultra-extreme case occurs when $9M^2=8Q^2=2/\L$, so that $P_+=P_-$.
Otherwise, the extreme black hole occurs if $M=M_-(Q,\L)$,
and the extreme naked singularity occurs if $M=M_+(Q,\L)$.
The generic black hole occurs if $M_-(Q,\L)<M<M_+(Q,\L)$,
and the generic naked singularity otherwise.

A constant-$T$ spatial surface of the generic naked-singularity solution
has topology $S^2\times R^1$,
which can be visualized as an $S^3$ punctured at opposite poles,
with the two singularities being at finite affine distance.
The electric field lines diverge from one pole and converge on the other,
so that the singularities have the appearance of point charges.

We now turn to the extreme cases.
The extreme black-hole solution can be interpreted as
a pair of oppositely charged extreme Reissner-Nordstr\"om black holes
at opposite poles of a de Sitter cosmos.
Note that for $\L>0$ the extreme case occurs for $M^2<Q^2$.
That is, in the cosmological context, the maximal charge
on a black hole---beyond which there is a naked singularity instead---is
larger than in asymptotically flat space \cite{RO}, cf.\ cite{BA}).
The topology of a constant-$T$ spatial surface is $S^2\times R^1$.
The two minimal-radius spheres, which also represent the black holes'
horizons, are located at infinite affine distance on this spatial surface,
preventing a wormhole identification, so that these black holes
are horns rather than wormholes.

The extreme naked-singularity case has a novel structure
that suggests the creation and subsequent annihilation
of a pair of point charges.
A constant-$T$ spatial surface has topology $S^2\times R^1$,
which again can be visualized as an $S^3$ punctured at opposite poles,
with the two singularities being at finite affine distance,
and the electric field lines
diverging from one pole and converging on the other.
Unlike the generic naked singularities,
or that of the negative-mass Schwarzschild solution,
these singularities do not exist for all time,
but develop from an initially regular state,
i.e.\ there are partial Cauchy surfaces,
of topology $S^2\times R^1$ or $S^2\times S^1$,
depending on the identifications made.
The singularities are not even weakly censored, in the sense that
any observer whose future life is long enough will see them:
any path from $\iota^-$ to $\iota^+$ passes through the causal future
(and the causal past) of the singularities.
This is in distinction to the generic black hole,
or the Reissner-Nordstr\"om black hole,
whose singularities remain unseen by wary observers.
Another novel feature is that the singularities subsequently dissolve,
with the process being visible from $\Scri^+$.

Finally, we note that the ultra-extreme case
is similar to the generic naked-singularity case,
except that the singularities are at infinite affine distance.

Another special case is of interest here (it can be generalized
to include more than two black holes), namely  $Q^2 = M^2$.
In the RN case this is the extremal black hole, but when $\Lambda \neq 0$
this choice does not force a double root --- the global structure and conformal
diagram of Fig.\ 6 still apply. What does change is the way the
cosmological coordinates cover the diagram: the
lens-shaped region degenerates into the line $\tau=0$, so that a
continuous region between the singularity and $\Scri^+$ is covered.
Therefore in this case the entire analytic extension of the metric
(6) can be obtained by gluing together alternate copies of expanding
and contracting cosmological coordinate patches. Figure 8 shows a
pair of such regions, with the contracting one outlined by thick lines.
(For easy comparison with Fig.\ 6 it may help to turn the latter
upside down.) We see that the expanding patch contains $\Scri^+$,
and the contracting one contains $\Scri^-$.
It is therefore not immediately clear,
if we calculate in contracting coordinates only,
how to identify the black hole event horizon as the boundary of the
past of $\Scri^+$. We can follow outgoing light rays to $r=\infty$,
but at that point their geometrical distance $R$, measured by the
Schwarzschild coordinate $R$ (or by the area of the sphere $r=$
const) is still finite, $R = R_c$. However, there is this difference
between such light rays and those that fall into the black hole, that
the latter reach the geometrical singularity, which {\em is} contained
in the contracting cosmological coordinates. Furthermore, timelike
geodesics heading toward the point S in the figure have an infinite
proper time. Thus S is a safe haven for observers who desire to avoid
the black hole, and it can be regarded as a small piece of $\Scri^+$ that
can be asymptotically reached in contracting cosmological coordinates,
just enough to be able to identify the event horizon.\footnote{This feature
holds for the generic case, but is particularly valuable for the
multi-black-hole solutions of Section 7.}

\bigskip

\unitlength=.75mm
\thinlines
\begin{picture}(141.00,133.00)(20,0)
\put(60.00,0.00){\line(-1,1){40.00}}
\put(20.00,40.00){\line(1,1){40.00}}
\put(60.00,80.00){\line(1,0){40.00}}
\put(80.00,81.00){\makebox(0,0)[cb]{$\Scri^+$}}
\put(80.00,40.00){\makebox(0,0)[ct]{$\Scri^-$}}
\thicklines
\put(100.00,80.00){\line(-1,-1){40.00}}
\put(60.00,40.00){\line(1,0){40.00}}
\put(100.00,40.00){\line(1,1){40.00}}
\put(140.00,80.00){\line(-1,1){40.00}}
\bezier{188}(60.00,40.00)(75.00,40.00)(100.00,60.00)
\bezier{188}(100.00,60.00)(125.00,80.00)(140.00,80.00)
\multiput(0,0)(0,-2){20}{
        \put(100.00,120.00){\makebox(0,0)[ct]{\bf*}}}
\thinlines
\bezier{312}(90.00,40.00)(120.00,80.00)(100.00,100.00)
\multiput(0,0)(0,-2){20}{
        \put(60.00,40.00){\makebox(0,0)[ct]{*}}}
\multiput(0,0)(2,-2){20}{
        \put(60.00,80.00){\makebox(0,0)[ct]{.}}}
\multiput(0,0)(2,-2){10}{
        \put(100.00,80.00){\makebox(0,0)[ct]{.}}}
\put(140.00,65.00){\vector(-1,0){25.00}}
\put(141.00,65.00){\makebox(0,0)[lc]{event horizon}}
\put(120.00,43.00){\vector(-1,0){22.00}}
\put(121.00,43.00){\makebox(0,0)[lc]{past de Sitter horizon}}
\put(129.00,110.00){\vector(-4,-1){15.00}}
\put(130.00,110.00){\makebox(0,0)[lc]{$r=0$}}
\put(130.00,99.00){\vector(-4,-1){24.00}}
\put(131.00,99.00){\makebox(0,0)[lc]{$r=$ const}}
\put(50.00,50.00){\makebox(0,0)[cc]{expanding}}
\put(122.00,82.00){\makebox(0,0)[cc]{contracting}}
\put(140.00,73.00){\vector(-3,1){12.00}}
\put(141.00,73.00){\makebox(0,0)[lc]{$\tau=$ const$<0$}}
\put(90.00,90.00){\vector(1,-1){9.00}}
\put(89.00,91.00){\makebox(0,0)[rb]{S}}
\put(46.00,73.00){\vector(2,-1){31.00}}
\put(45.00,73.00){\makebox(0,0)[rb]{$r=\infty$}}
\put(32.00,63.00){\vector(2,-1){7.00}}
\put(31.00,63.00){\makebox(0,0)[rc]{$r=0$}}
\end{picture}

\vspace{.5truecm}
{\noindent \small
{\bf Fig.\ 8.} Two patches of cosmological coordinates for the RNdS
geometry for the case $Q^2=M^2$ and $p<1$ (``undermassive'').}

\bigskip\bigskip

The $Q^2=M^2$ RNdS geometries still depend on two parameters, $M$
and $\Lambda$ resp.\ $H$.
A generic black hole occurs if $\L M^2<3/16$,
an extreme naked singularity if $\L M^2=3/16$,
and a generic naked singularity if $\L M^2>3/16$.
The extreme black hole does not occur in this class.
If $\L M^2 \ll 1$
the Killing horizons are located approximately at
$$R_i\sim M-HM^2,\qquad R_o\sim M+HM^2,\qquad R_c\sim H^{-1}-M.$$
(for higher order of approximation see Romans \cite {RO}.)
Stated in terms of the dimensionless parameter $p=4M|H|$, if
$p<1$ we have the generic black hole or ``undermassive'' case.
If $p>1$ there is only one real root of $F(R)=0$. In this
generic nakedly singular or ``overmassive'' case
the outer black hole horizon and the
de Sitter horizon have disappeared, only what used to be
the inner black hole horizon remains. One could also interpret the
remaining horizon as a cosmological one, separating the two naked
singularities at antipodal regions of a background de Sitter space.
The conformal diagram for this case, constructed according to the
block gluing rules, is shown in Fig.\ 9.

\unitlength=.75mm
\begin{picture}(106.00,60)(22,0)
\thinlines
\multiput(0,0)(0,-2){20}{
\put(100.00,45.00){\makebox(0,0)[ct]{*}}}
\multiput(0,0)(2,2){20}{
\put(60.00,5.00){\makebox(0,0)[ct]{.}}}
\put(60.00,45.00){\line(1,0){40.00}}
\put(80.00,46.00){\makebox(0,0)[cb]{$\Scri^+$}}
\bezier{128}(60.00,30.00)(76.00,16.00)(76.00,5.00)
\put(80.00,4.00){\makebox(0,0)[ct]{$\Scri^-$}}
\put(105.00,20.00){\vector(-2,-1){20.00}}
\put(106.00,20.00){\makebox(0,0)[lc]{$\tau=$ const $<0$}}
\put(105.00,30.00){\vector(-2,-1){26.00}}
\put(106.00,30.00){\makebox(0,0)[lc]{$\tau=$ const $>0$}}
\put(50.00,24.00){\vector(1,0){15.00}}
\put(49.00,24.00){\makebox(0,0)[rc]{$r=$ const}}
\put(71.50,13.00){\makebox(0,0)[cc]{\small contracting}}
\put(85.00,35.00){\makebox(0,0)[cc]{\small expanding}}
\thicklines
\put(60.00,5.00){\line(1,0){40.00}}
\put(100.00,5.00){\line(-1,1){40.00}}
\bezier{176}(60.00,5.00)(79.00,14.00)(100.00,5.00)
\bezier{176}(60.00,20.00)(80.00,19.00)(100.00,5.00)
\multiput(0,0)(0,-2){20}{
\put(60.00,45.00){\makebox(0,0)[ct]{\bf*}}}

\end{picture}

\vspace{.6truecm}
{\noindent \small
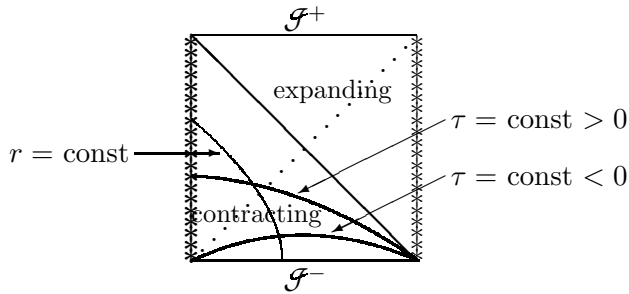
{\bf Fig.\ 9.} Two patches of cosmological coordinates for the RNdS
geometry for the case $Q^2=M^2$ and $p>1$ (``overmassive''). The thick
outline shows the contracting patch.}

\bigskip \bigskip \noindent
{\bf 4.2 Collapsing Dust}

\bigskip \noindent
The solutions discussed so far correspond to ``eternal'' black holes (or a
combination of white holes and black holes). The future part of the usual black
hole geometry can be generated from matter initial data, say by the collapse
of a sphere of dust. A similar process is possible for $Q^2=M^2$ RNdS.
It is of interest here because the dust can ``cover up'' an initial black
hole singularity, and will be used for that purpose in section 7.

To generate such a geometry from dust, the dust must itself have $Q^2=M^2$,
so that electrical forces largely balance gravitational forces. The dust can
still
collapse if it has the right initial velocity. A simple situation is dust that
is at
rest in cosmological coordinates: in any metric and potential of form (6), for
an arbitrary $U$, such dust remains at rest ($r=$ const), if considered as
test particles. A typical wordline is shown in Fig.\ 8. To understand why
it behaves this way we look more closely at the electric field associated
with this geometry. We know that the parameter $Q$ of the black hole on
the left is opposite to that of the one on the right. Suppose the left
hole is negative and the right one positive. On the spacelike surface
corresponding to a horizontal line drawn through the center of
Fig.\ 8, the electric field will then point from right to left. By flux
conservation, the electric field will therefore also point to the left
in the region near the upper singularity shown in the Figure. Thus we
can associate a {\em negative} charge with this singularity. (The
singularity to the right of the $r=0$ inner horizon, which is not shown
in the Figure, correspondingly has positive charge). The dust particle
on the $r=$ const trajectory has the same charge as the black hole that
is included in that coordinate patch, namely positive. It therefore ends
up on the negative singularity.

Any point in Fig.\ 8 can be considered to be in either of two possible
cosmological coordinate patches. For example, a point near $\Scri^-$ is
in the contracting patch shown, which contracts about the right black
hole. By reflecting this patch about a vertical line through the center
of the Figure we obtain a patch that contracts about the left black hole.
It also contains the region near $\Scri^-$. Its radial coordinate will be
denoted by $r'$. The trajectory $r'=$ const, obtained by reflecting the
$r=$ const trajectory shown in the Figure, describes a negative test charge
that falls into the positive singularity of the left (negative) black hole.
Similarly the region between the de Sitter and the event horizon in the
contracting patch has trajectories of positive charges that fall into the
positive black hole, and of negative charges that go to $\Scri^+$. The
region inside the event horizon has trajectories that go to one or the
other singularity, depending on their charges. Of course all these
trajectories that are simply described by $r=$ const or $r'=$ const
satisfy special initial conditions --- their initial position is arbitrary,
but their initial velocity is then determined.

To take into account the effect
of the dust {\em on} the metric and potential, one needs to match the vacuum
region to an interior solution. For spherical symmetry the matching conditions
are equivalent to the demand that the dust at the boundary of the interior
region
move on a test particle path of the vacuum region. Thus a possible boundary
for a collapsing (expanding) ball of dust is $r=$ const in collapsing
(expanding) cosmological coordinates. The surface area $4\pi r^2 U^2$ of the
dust ball collapses to zero when $U = 0$, which is also the location of
the geometrical singularity; at that point the center of the dust ball must
coincide with the surface.  The corresponding conformal diagram therefore
looks as shown by the thick curves in Fig.\ 10. The region filled with dots
denotes the location of the dust.

\unitlength=2.0mm
\begin{picture}(40.00,28.00)(0,5)
\thicklines
\put(10.00,10.00){\line(1,0){13.00}}
\put(20.00,20.00){\line(-1,-1){10.00}}
\bezier{20}(25.00,18.33)(26.50,20.00)(25.00,22.00)
\put(18.20,10.00){\line(5,6){6.92}}
\put(20.00,28.00){\line(5,-6){5.08}}
\bezier{44}(25.00,23.00)(29.00,19.00)(26.00,15.00)
\put(20.00,28.00){\line(1,-1){5.00}}
\put(26.08,15.00){\line(-3,-5){3.00}}
\thinlines
\put(30.00,28.00){\line(-5,-6){5.00}}
\put(32.00,10.00){\line(-5,6){7.08}}
\bezier{30}(25.00,18.33)(23.17,20.00)(25.00,22.00)
\put(40.00,10.00){\line(-1,1){10.00}}
\put(30.00,28.00){\line(-1,-1){5.00}}
\bezier{60}(25.00,23.00)(21.00,19.00)(24.00,15.00)
\put(24.00,15.00){\line(3,-5){3.00}}
\put(27.00,10.00){\line(1,0){13.00}}
\multiput(0,0)(0,0.75){11}{
\put(20.00,20.00){\makebox(0,0)[cc]{\bf *}}}
\multiput(0,0)(.8,-.8){7}{
\put(20.60,19.40){\makebox(0,0)[cb]{\bf.}}}
\multiput(0,0)(0,0.75){11}{
\put(30.00,20.00){\makebox(0,0)[cc]{*}}}
\put(20.00,11.00){\circle*{0.00}}
\put(21.00,11.00){\circle*{0.00}}
\put(22.00,11.00){\circle*{0.00}}
\put(23.00,11.00){\circle*{0.00}}
\put(24.00,12.00){\circle*{0.00}}
\put(23.00,12.00){\circle*{0.00}}
\put(22.00,12.00){\circle*{0.00}}
\put(21.00,12.00){\circle*{0.00}}
\put(21.00,13.00){\circle*{0.00}}
\put(22.00,13.00){\circle*{0.00}}
\put(23.00,13.00){\circle*{0.00}}
\put(24.00,13.00){\circle*{0.00}}
\put(25.00,14.00){\circle*{0.00}}
\put(24.00,14.00){\circle*{0.00}}
\put(23.00,14.00){\circle*{0.00}}
\put(22.00,14.00){\circle*{0.00}}
\put(23.00,15.00){\circle*{0.00}}
\put(25.00,15.00){\circle*{0.00}}
\put(26.00,16.00){\circle*{0.00}}
\put(25.00,16.00){\circle*{0.00}}
\put(24.00,16.00){\circle*{0.00}}
\put(25.00,17.00){\circle*{0.00}}
\put(27.00,18.00){\circle*{0.00}}
\put(26.00,18.00){\circle*{0.00}}
\put(26.00,19.00){\circle*{0.00}}
\put(27.00,19.00){\circle*{0.00}}
\put(26.00,20.00){\circle*{0.00}}
\put(27.00,20.00){\circle*{0.00}}
\put(26.00,21.00){\circle*{0.00}}
\put(25.50,22.00){\circle*{0.00}}
\put(24.92,22.67){\circle*{0.00}}
\put(24.33,23.25){\circle*{0.00}}
\put(23.75,23.92){\circle*{0.00}}
\put(23.17,24.58){\circle*{0.00}}
\put(22.50,25.25){\circle*{0.00}}
\put(26.00,17.00){\circle*{0.00}}
\put(32.00,24.00){\vector(-2,-1){5.50}}
\put(32.25,24.00){\makebox(0,0)[lc]{$r=0$}}
\put(18.00,24.00){\vector(2,-1){5.33}}
\put(17.75,24.00){\makebox(0,0)[rc]{$r'=0$}}
\put(14.00,18.00){\vector(1,-1){2.00}}
\put(14.00,18.00){\makebox(0,0)[rc]{$r=\infty$}}
\put(36.25,18.25){\makebox(0,0)[lc]{$r'=\infty$}}
\put(36.00,18.00){\vector(-1,-1){2.00}}
\end{picture}

\vspace{-.6truecm}
{\noindent \small
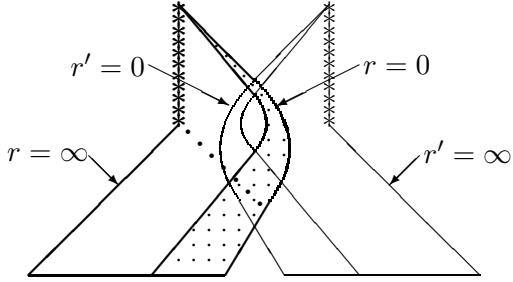
{\bf Fig.\ 10.} Conformal diagram for a RNdS black hole generated by the
collapse of a ball of charged dust in a de Sitter background (thick
lines). The dust region is shown dotted. The dotted line is the black hole
event horizon. The thin lines show a dust ball that is symmetrically
placed at the antipodal region of the RNdS background, and has
opposite charge. If both dust balls are present only the region between
the curves $r=0$ resp.\ $r'=0$ applies.
}

\bigskip\bigskip
Because the dust includes the origin $r=0$, there is now no continuation
to the right of the heavily outlined region necessary or possible. There
is still a cosmological horizon, $r=\infty$, and the continuation on the
other side could be analytic (a semi-infinite ``string of beads'') or
reflection-symmetric about a vertical axis through $\Scri^-$ (a collapsing
dust ball at the antipodal region). Of the two geometrical singularities
associated with one RNdS black hole, the dust covers up only the one that
would have been on the right in the figure, with the same total charge as
that of the dust. The other singularity could be covered up by another
dust ball, with the opposite charge. This is shown by the thin curves
in Figure 10. If both dust spheres are present, then the physical part of
the geometry lies between the $r'=0$ curve on the left and the $r=0$ curve
on the right. If conditions are as shown, they uncover and then cover up
again a (small) region of vacuum RNdS geometry between them.

\bigskip \bigskip
\noindent
{\large\bf 5 Euclidean Metrics}

\bigskip \noindent
Complex analytic continuations of a spacetime metric can change the metric's
signature. The new metric will solve the same (analytic) equations
(e.g. sourceless Einstein, Einstein-Maxwell, etc) as the original metric.
This is most frequently used to obtain positive definite or ``Euclidean''
solutions of the Einstein equations. Such a continuation will embed the
original spacetime in a higher-dimensional space with a complex metric,
and the change to a Euclidean metric can be thought of as a rotation
from the original, Lorentzian section to the final, Euclidean one.
At fixed points the rotation acts also as a rotation on the tangent
spaces, and it is clear that the desired change in signature typically
leaves a 3-dimensional (real) subspace invariant. Therefore two sections with
a real metric typically intersect in a 3-dimensional hypersurface. This
hypersurface will be time-symmetric. In time orthogonal coordinates
the metric coefficients will therefore be even functions of the time
coordinate, and the continuation is simply described by the familiar
replacement
$t \rightarrow it$.

Thus any time-symmetric analytic geometry has a Euclidean continuation.
The best known Euclidean metrics are obtained from the even simpler
case when there is not only the discrete time-symmetry, but also a
timelike Killing vector, which must then be orthogonal to the surface of
time symmetry; in other words, the original spacetime is static.  We consider
a few of these cases with particular regard to their global structure.

The Euclidean partner of Minkowski space is of course flat Euclidean space:
rectangular coordinates $x,\, y,\, z,\, t$ are a time-symmetric description,
and hence suitable for this extension via $t \rightarrow it. $\footnote{It is
amusing to contemplate that today we can describe Euclidean space via
``imaginary time" in Minkowski space, whereas Minkowski was describing
his space by an imaginary coordinate of Euclidean space!}  One can think up
many other coordinates in which the replacement works, with the same
result (but with different physical interpretation). One is the ``Rindler
frame'', with the metric
\begin{equation}
ds^2 = - x^2 dt^2 + dx^2 + dy^2 + dz^2 \qquad (x \geq 0).
\end{equation}
Here the replacement $t \rightarrow it$ yields Euclidean space in cylindrical
coordinates, provided that the Euclidean $t$-coordinate, which plays the role
of the angle variable, has the correct periodicity $2 \pi$. (If the periodicity
were not right, the Euclidean space would not be regular at the origin, but
have a conical singularity there.)

A periodicity is forced on the Euclidean ``time" coordinate whenever the
Lorentzian metric can locally be put in the form of Eq (8). This will
typically happen along a (2-dimensional) ``axis,'' but there may be more
than one such axis. In that case the Lorentzian metric can be regular
on both axes, but not the Euclidean metric --- unless the periodicities
of $t$ required for regularity happen to agree at both places. We will
see an example of this in de Sitter space.

The Schwarzschild geometry, being static, also has a Euclidean partner.
Because it takes the form of Eq (8) near the horizon, the Euclidean time
must be periodic. This is easily seen in the metric's isotropic form,
$$ ds^2 = -\left({r-M/2 \over r+M/2}\right)^2 dT^2 +
\left(1+{M\over 2r}\right)^4\,\left(dr^2 + r^2\,d\Omega^2\right).$$
Near the horizon, $r = M/2$, this has the form (8) with
$4(r - {M\over 2}) = x$, $t=T/4M$, and $16M^2\,d\Omega^2 \sim dy^2 + dz^2$.
In the Euclidean version, then, $t$ must have periodicity $2\pi$, hence
$T$ will have periodicity $8\pi M$.

Because Euclidean space has no conformally invariant lightcones and
no $\Scri^\pm$, a conformal representation of such spaces is not as revealing
as an embedding in a flat higher-dimensional space. Two-dimensional
subspaces of interest can often be embedded in flat 3-dimensional space.
This is the case for the $r,\, t$ section of the Euclidean Schwarzschild
geometry. In Schwarzschild coordinates embedded in flat space cylindrical
coordinates (with $T$ the angle) we have
$$ \left(1-{2M\over R}\right)\,dT^2 + \left(1-{2M\over R}\right)^{-1}\,dR^2
= \rho^2 dT^2 + d\rho^2 + dx^2$$
hence $\rho^2  = \left(1-{2M\over R}\right)$ and $dx^2 =
\left((1-\rho^2)^{-4} -1\right)d\rho^2$ specifies the curve whose figure
of rotation gives the embedded surface. (Here we have put $4M=1$ to
achieve the proper periodicity, which means that the figure is smooth
at its tip.) This embedding is shown in Fig.\ 11.

\newpage

\bigskip
\unitlength 0.150mm
\begin{picture}(363.33,140.00)(-200,0)
\thicklines
\bezier{60}(64.17,64.17)(4.17,45.00)(0.00,0.00)
\bezier{150}(64.76,64.67)(136.76,88.67)(344.76,90.00)
\bezier{60}(64.17,-64.17)(4.17,-45.00)(0.00,0.00)
\bezier{150}(64.76,-64.67)(136.76,-88.67)(344.76,-90.00)
\bezier{40}(329.66,57.57)(344.74,123.17)(359.83,57.57)
\bezier{40}(359.83,57.57)(367.37,-3.53)(359.83,-55.57)
\bezier{40}(359.83,-55.57)(344.74,-127.21)(329.66,-55.57)
\bezier{50}(329.66,-55.57)(317.08,2.51)(329.66,57.57)
\thinlines
\put(0.00,0.00){\vector(0,1){100.00}}
\put(0.00,105.00){\makebox(0,0)[cb]{$\rho$}}
\put(385.00,0.00){\makebox(0,0)[lc]{$x$}}
\bezier{55}(160.00,60.00)(150.00,30.00)(150.00,-25.00)
\put(150.00,-20.00){\vector(0,-1){5.00}}
\put(150.00,-30.00){\makebox(0,0)[ct]{$T$}}
\put(0.00,0.00){\line(1,0){20.00}}
\put(40.00,0.00){\line(1,0){20.00}}
\put(80.00,0.00){\line(1,0){20.00}}
\put(120.00,0.00){\line(1,0){20.00}}
\put(160.00,0.00){\line(1,0){20.00}}
\put(200.00,0.00){\line(1,0){20.00}}
\put(240.00,0.00){\line(1,0){20.00}}
\put(280.00,0.00){\line(1,0){20.00}}
\put(325.00,0.00){\vector(1,0){55.00}}
\end{picture}

\vskip.7in

\noindent
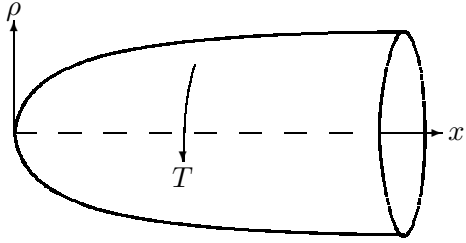
{\small {\bf Fig.\ 11.} Embedding of the $R,\, T$ section of the Euclidean
Schwarzschild solution in flat 3-dimensional space.}
\bigskip \bigskip

We see that the {\em topology} of this surface is $\rm I\!\!R^2$, and that
of the 4-dimensional space is $\rm I\!\!R^2 \times S^2$. However,
asymptotically the geometry is cylindrical ($\rm I\!\!R^1 \times S^1\times
S^2$). Other asymptotically flat Euclidean black hole
metrics, obtained e.g. from $t \rightarrow it\,$ in the RN geometry, result
in similar figures. In each case the Euclidean metric extends down only to
the $r$-value corresponding to the outermost horizon, and the horizon itself
is the single point on the axis of rotation.
One might ask, how can electric field lines point radially outward at
infinity without being singular somewhere in Fig.\ 11? Here it helps to
recall that the electric field is a 2-form, it is proportional to the
area 2-form of the Figure. As the only boundary of that area is the
circle drawn at large $x$, so there can be electric flux at infinity
without another surface for that flux to come from. Similarly, a magnetic
charge would be proportional to the area form of the spherical, orthogonal
space (which is not shown in Fig.\ 11).

The embedding looks different for the extreme cases
(e.g.\ $Q^2 = M^2$): the ``point on the axis'' is infinitely
far away, so the surface comes to an infinitely long ``spike'' of
ever-decreasing radius, {\em if} Euclidean time is given periodic
identification (Fig.\ 12). But because the point on the axis is never
reached, there is no conical
singularity to be avoided, and no unique period necessary for regularity.
In fact, the Euclidean $T$ can equally well have infinite range in that
case.\footnote{It may be that not all of these are quantum mechanically
stable: in the corresponding flat space case, any finite period (``hot
flat space'') is unstable.}

\bigskip

\unitlength .27mm
\begin{picture}(206.00,64.79)(-200,0)
\thicklines
\bezier{70}(-96.00,0.01)(-34.00,0.01)(-9.33,21.83)
\bezier{200}(-9.33,21.83)(22.67,49.30)(193.33,47.68)
\bezier{70}(-96.00,-1.61)(-34.00,-1.61)(-9.33,-23.43)
\bezier{200}(-9.33,-23.43)(22.67,-50.90)(193.33,-49.28)
\bezier{40}(186.00,30.00)(194.00,64.79)(202.00,30.00)
\bezier{40}(202.00,30.00)(206.00,-2.40)(202.00,-30.00)
\bezier{40}(202.00,-30.00)(194.00,-67.99)(186.00,-30.00)
\bezier{50}(186.00,-30.00)(179.33,0.80)(186.00,30.00)
\bezier{5}(-96.08,-1.40)(-97.00,-1.00)(-96.17,-0.00)
\end{picture}

\bigskip \bigskip
\bigskip \bigskip

\noindent
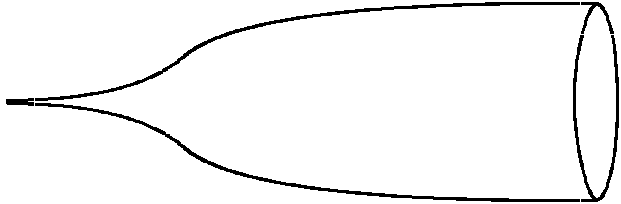
{\small {\bf Fig.\ 12.} Embedding of the $R,\, T$ section of the Euclidean
extremal Reissner-Nordstr\"om solution in flat 3-dimensional space. The
coordinates are similar to those in Fig.\ 11. The embedding is given by
$dx^2 = \left(\rho^{-2}(1-\rho)^{-4}-1\right)d\rho^2$. The periodicity of
$T$ was assumed to be $2\pi$, but this is not necessary. $T$ can have any
periodicity, or no periodicity at all. The latter case corresponds to the
universal covering space of the Figure.}
\bigskip \bigskip

The Euclidean continuation of de Sitter space is quite simple. Since the
curvature is positive constant, the Euclidean partner is the round 4-sphere.
One can verify this e.g.\ in the static frame or in the usual `hyperbolic'
de Sitter coordinates. The Euclidean static frame describes the $R,\,T$
subspace (which is a round 2-sphere) in terms of equal-area
coordinates.\footnote{These are obtained by projecting the 2-sphere
(when embedded in 3-space)
at constant latitude onto a cylinder that is tangent to the 2-sphere at
its equator.} In hyperbolic de Sitter coordinates, the line element is
$$ds^2 = 3\Lambda^{-1}(-d\theta^2 + \cosh^2\theta\,d\bigcirc^2),$$
where $d\bigcirc^2$ denotes the line element on the round 3-sphere.
When $\theta \rightarrow i\theta$ these coordinates become the
usual Euclidean polar coordinates on the 4-sphere
(except that $\theta$ is measured from the equator rather than from the
pole). In either case the Euclideanized coordinate ($T$ resp.\ $\theta$)
must have a well-defined period to make the continuation nonsingular.
The cosmological coordinates (Eq 5) are not time-symmetric and therefore
do not yield a Euclidean section by $t \rightarrow it$, unless we also
change $H$ into $iH$. This implies $\Lambda \rightarrow -\Lambda$,
so we would get the Euclidean partner of anti-de Sitter space.

The static part of the RNdS metric (between the outer black hole horizon
$R_o$ and the cosmological horizon $R_c$) can also be continued to
Euclidean time. Each horizon requires a periodicity. To find this for
a metric of type (1), we expand $F(R)$ around one of its zeros
(where $R$ has the horizon value $R_H$), letting $R-R_H \equiv z^2$:
$$ F(R) = F'(R_H) z^2 + \dots \qquad \qquad dR^2 = 4z^2(dz)^2$$
so that the metric becomes
$$ds^2 \approx F'z^2dT^2 + {4 dz^2\over F'} + R_z^2d\Omega^2 =
{4\over F'}\left(z^2d(F'T/2)^2 + dz^2\right) + R_z^2d\Omega^2.$$
Therefore $z,\,F'T/2$ are polar coordinates regular at the origin if
$|F'|T/2$ has period $2\pi$, or $T$ has period $4\pi/|F'(R_H)|$.
For Schwarzschild this yields the well-known periodicity $T_0=8\pi M$, as we
already found above.
For the RNdS solution one finds that the periods required by the two
horizons are equal when $Q^2=M^2$; the period then has the value
$T_0= 2\pi/H\sqrt{1-4MH}$.

There is another special case of RNdS where one can have equality of periods.
We saw that the extremal horizons do not require any particular periodicity
when Euclideanized. Hence when the inner and outer black hole horizons
coincide (see Section 4.1), we can simply assign them the periodicity
of the cosmological horizon and thus obtain a regular Euclidean geometry.

An embedding of the $R,\,T$ subspace of the non-extremal Euclidean RNdS space
looks similar to
Fig.\ 11, except that instead of being asymptotically cylindrical, the surface
is closed off on the right by a nearly hemispherical cap whose center is
the cosmological horizon; the topology of the 4-dimensional space is
therefore $\rm S^2 \times S^2$. The extremal case would be shown by
Fig.\ 12 similarly modified.

Flat 3D space cannot, of course, portray more than these two-dimensional
subspaces (and the analogous $r, \phi$ subspaces that are the same as
in the Lorentzian geometries). To show, say, the $\phi$-direction and
fix only $\theta$ at $\pi/2$ one would need another independent axis of
rotation. But the $x$-coordinate in Fig.\ 11 does not differ by more than
a factor of 2 from $R - 2M$. We can therefore get an approximate idea
of the 3-geometry at the fixed $\theta$ by rotating the figure about a
vertical axis at distance $2M$ from $x=0$. Each point in 3-space between
the vertical limits of the figure is covered twice (by the front and by
the back of the surface) except the points swept out by the vertical
section of the figure. Thus one obtains two copies of this region, sown
together at their boundaries. The $T =$ const line on Fig.\ 11 that
represents its section by a vertical plane sweeps out the most geometrically
accurate rendition of the $R,\,\phi$ subspace: it has the familiar
``wormhole" shape.\footnote{It is not quite geometrically accurate, because
Fig.\ 11 is not a paraboloid, whereas the $R,\,\phi$ subspace is generated
by rotating the Flamm parabola.} If one confines attention to half the
space, which is covered once during the rotation, say be the front half
of the surface in Fig.\ 11, then all the $T =$ const, $R,\,\phi$ subspaces
seem to end at the horizon, except the one generated by the vertical cut.
The other surfaces of course do not stop either, but can be smoothly
continued into the other half of the space (generated by the back part of
the surface).

The same construction applied to RNdS yields the half space shown in Fig.\ 13.
Again the geometry of the outermost $T =$ const surface is fairly accurately
represented. It is the same geometry as that of a $T =$ const surface in
Lorentzian RNdS, namely Fig.\ 7. (You have to rotate Fig.\ 7 by 90$^o$
and flip the two funnels to the inside so the circles $R=R_o$ can be
identified.)

\bigskip
\unitlength 1.00mm
\linethickness{0.4pt}
\begin{picture}(90.67,40.00)
\thicklines
\bezier{216}(50.00,37.00)(29.33,20.00)(50.00,2.67)
\thinlines
\bezier{104}(50.00,2.67)(56.33,0.67)(56.00,20.00)
\bezier{104}(50.00,37.33)(56.33,39.33)(56.00,20.00)
\thicklines
\bezier{216}(70.00,37.00)(90.67,20.00)(70.00,2.67)
\thinlines
\bezier{104}(70.00,2.67)(63.67,0.67)(64.00,20.00)
\bezier{104}(70.00,37.33)(63.67,39.33)(64.00,20.00)
\bezier{108}(46.17,22.17)(32.83,20.00)(46.33,17.83)
\bezier{112}(46.33,17.83)(60.00,15.50)(73.83,17.83)
\bezier{104}(73.83,17.83)(86.83,20.00)(73.83,22.17)
\bezier{112}(73.83,22.17)(59.83,24.17)(46.00,22.17)
\bezier{76}(52.00,37.33)(60.17,32.33)(67.67,37.33)
\thicklines
\bezier{80}(50.17,37.33)(60.50,40.00)(69.67,37.33)
\bezier{84}(50.50,2.50)(60.33,-2.17)(69.67,2.50)
\thinlines
\bezier{36}(56.00,19.83)(60.17,18.17)(63.83,19.83)
\bezier{36}(63.83,19.83)(60.17,21.67)(56.00,20.00)
\bezier{140}(39.67,20.00)(50.17,4.67)(56.00,20.00)
\bezier{140}(80.33,20.00)(69.83,4.67)(64.00,20.00)
\bezier{80}(51.00,11.06)(60.22,7.28)(69.00,11.06)
\bezier{44}(69.00,11.06)(73.33,12.94)(67.56,13.94)
\bezier{36}(51.22,11.06)(47.44,12.83)(51.78,13.72)
\put(35.78,7.06){\vector(4,1){8.67}}
\put(35.33,6.94){\makebox(0,0)[rc]{0}}
\put(36.00,13.06){\vector(3,1){8.00}}
\put(35.33,12.94){\makebox(0,0)[rc]{$T_0/8$}}
\put(35.78,19.72){\vector(1,0){5.33}}
\put(35.22,19.72){\makebox(0,0)[rc]{$T_0/4$}}
\put(36.00,28.06){\vector(4,-1){6.67}}
\put(35.22,28.06){\makebox(0,0)[rc]{$T_0/2$}}
\put(83.89,24.50){\vector(-3,-4){3.33}}
\put(84.22,24.39){\makebox(0,0)[lc]{$R_c$}}
\put(69.33,25.39){\vector(-1,-1){5.22}}
\put(69.78,25.28){\makebox(0,0)[lc]{$R_o$}}
\end{picture}

\bigskip
\noindent
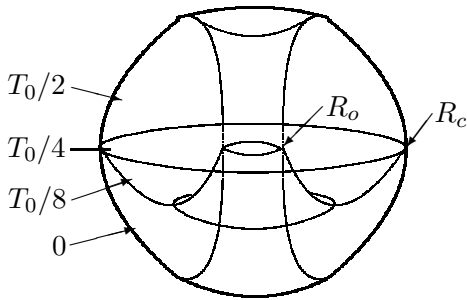
{\small {\bf Fig.\ 13.} Half of the 3D subspace $\theta = \pi/2$ of Euclidean
RNdS looks like a cored apple. To get the complete subspace, imagine
another cored apple identified with the first one everywhere along its
surface (including the cored ``tunnel'').
Here $Q^2=M^2$ so that the space is regular at both horizons. The
cosmological horizon is the outer ``equator,'' and the black hole horizon
is the ``inner equator.'' The outer surface of the figure plus the
tunnel surface corresponds to constant values of $T$ ($T=0$ and
$T=T_0/2$). Two other surfaces of constant $T$ are shown, cutting
through the interior of the cored apple, and
intersecting each other at the horizons. The intrinsic geometry of all these
surfaces is the same, but only the one at $T=0$ gives a somewhat accurate
rendition of its intrinsic geometry.}

\bigskip \bigskip

The same method, using a surface of symmetry, allows us to go from a
Euclidean solution back to a Lorentzian one. If the Euclidean space has
several symmetries, we do not necessarily have to go back to the
Lorentzian space we may have started from. This happens, for example, if
we make a Lorentzian continuation ``across" the equatorial symmetry
plane of the Euclidean Schwarzschild solution. $T$ then remains spacelike
(and periodic), but $\theta$ becomes timelike. The result is the metric
mentioned in Section 2. Although this metric has timelike Killing vectors
in any region (just as there is a Killing vector on the sphere that
agrees with $\partial/\partial \theta$ at any one point except the poles),
it is globally dynamical --- it resembles the de Sitter universe in that
respect. So one Euclidean solution can be the partner of several Lorentzian
ones.

\bigskip \bigskip
\noindent
{\large\bf 6 Physical Interpretation of Euclidean Solutions,

\hskip-2mm and a remark about the Gravitational Action}

\bigskip
\noindent
Now that we know all about Euclidean solutions of the Einstein equations,
what are they good for? We are familiar with analytic extensions into the
complex for convenience of calculation or to define quantities that are
otherwise not well defined. Similarly, Euclidean spacetimes can be regarded
as an elaborate ``change of contour.'' Such a change is appropriate in
connection with thermal states and with tunneling states, and these are
indeed the two important ways of using Euclidean geometries.
One can take a point of view that does not regard
these these two ways as different, but frequently the distinction makes
sense. For example, as a continuation of the ordinary Schwarzschild solution,
the Euclidean Schwarzschild solution is the arena for thermal states;
as a continuation of the dynamical metric of Section 2 it is most
reasonable as a tunneling state. On the other hand, the Euclidean
continuation of the RNdS solution can have either interpretation.
There are interesting unsolved question connected with these interpretations.
For example, in the thermal context there are questions about the
source of black hole entropy, the entropy value of extremal black holes,
and the ``information puzzle.'' In the realm of tunneling one would like
a better understanding of the boundary conditions, and of the theory that
is being approximated. I will not address these unsolved problems, but
confine attention to a few simple examples.

\bigskip\bigskip
\noindent
{\bf 6.1 Thermal Interpretation}

\bigskip\noindent
The connection between periodicity in imaginary time and finite temperature
is well known \cite{GH}. When the background is periodic Euclidean with
period $T_0$, only states with (inverse) temperature $\beta =\hbar/k{\cal T}
= T_0$ have finite stress-energy. Any field in the Schwarzschild background
at equilibrium should therefore acquire the temperature $\beta = 8\pi M$.
This agrees with the Hawking temperature at which the black hole itself
radiates. Similarly the periodicity of imaginary Rindler time (Eq 8)
corresponds to the temperature seen by accelerated observers.

In a WKB-type approximation to the sum over histories expression for the
partition function the most important contribution comes from the classical
(Euclidean) action $I$. When the Euclidean space has a finite size in the
$T$-direction the action of the background space itself is finite and
contributes to the free energy and entropy.\footnote{Existence of
gravitational and other degrees of freedom apparently
does not change this value of the
black hole entropy. This is the ``species problem.'' The explanation may be
that existence of additional species of field changes the renormalization
of the gravitational constant to compensate.} What is the Euclidean
black hole action? The Hilbert action, ${1\over 16\pi}\int {\cal R} dV$,
vanishes for any solution of the sourceless Einstein equations (because
they imply ${\cal R}=0$, where $\cal R$ is the curvature scalar). But in
order that the action be properly additive in the path integral one must
add a surface term to the Hilbert action. The surface term is
$-{1\over 8\pi}\int K d\Sigma$, where $\Sigma$ is the surface and $K$
the trace of its extrinsic curvature. With this term
the action has the opposite problem from before, it is infinite (because
$K$ falls off only as $1/r$ even in flat space, but $d\Sigma$ increases
as $r^2$). One therefore subtracts the flat space contribution,
${1\over 8\pi}\int K_0 d\Sigma$, where $K_0$ is the extrinsic curvature
of $\Sigma$ when embedded in flat space.\footnote{Not every $\Sigma$
can be embedded in flat space, the definition really works only for
asymptotically flat (or deSitter, see \cite{HH}) spaces in the limit
of $\Sigma$  becoming asymptotic as well. The corresponding prescription
in Lorentzian spacetimes works only for boundaries that consist of
spacelike and timelike parts meeting orthogonally, and the correction is
to be applied only to the timelike parts of the boundary. We do not know
any less ad-hoc definition of the action that still gives a finite
black hole action for finite (Lorentzian) times.}

For the Euclidean Schwarzschild black hole this action is easily computed:
at large $R$ we have
\begin{equation}
K = {2\over R} - {M\over R^2}, \qquad K_0 = {2\over R} \qquad {\rm so}
\quad I = -{1\over 8\pi}\int_{T=0}^{8\pi M}(K-K_0)\,d\Sigma = 4\pi M^2.
\end{equation}
If $\exp(-I/\hbar)$ is the main contribution to the partition function
$Z=\exp(-\beta F)$, we have $F = I/\hbar\beta$. Setting $\hbar=1=k$
(in addition to $G=1=c$ as we have all along) to
express all quantities in Planck units we find the value of the free energy
$F = {1\over 2}M$. If the black hole had no entropy we would expect $F=M$,
so the contribution of the entropy is $-{\cal T}S = -{1\over 2}M$, and
since ${\cal T} = 1/\beta =1/8\pi M$ we find $S=4\pi M^2$. This is usually
attributed to the horizon. How can we associate free energy and action
with the Euclidean horizon, which is just one point in the $R,\,T$ surface?

Several ways have been proposed. To find
$S = \beta^2\partial F/\partial\beta$
we need to know how the action changes with $\beta\, (=T)$.
If we change $\beta$ by changing the periodicity but not the local
geometry outside the horizon, we would introduce a conical singularity at
the horizon. The tip of the cone can be considered to have a
$\delta$-function scalar curvature $\cal R$. Its contribution to the
Hilbert action gives the correct $\beta$-dependence to $I$. Alternatively
we can keep the surface regular, but consider how the action is built up
out of small steps in $\beta$ starting from $\beta = 0$ (Fig.\ 14).
A change by $d\beta$ changes the
action by that of a wedge of angle $d\alpha = 2\pi (d\beta/8\pi M)$.
Because now ${\cal R} = 0$ everywhere on this regular surface, the action
of the wedge is again due only to the boundary. The large parts of the
boundary have $K=0$, so there is no contribution from them. The short
boundary $d\beta$ at infinity contributes, as in Eq (9), ${1\over 2} M d\beta$.
But now there is also a contribution from the horizon because of the
sharp corner there --- a $\delta$-function in $K$. When $\beta$ and $\alpha$
have reached the value where the whole surface is covered ($\alpha=2\pi$),
all the finite boundaries must cancel. For an arbitrary $\beta$ the horizon
contribution is therefore
$$-{1\over 8\pi}\int K d\Sigma = {1\over 8\pi}\left((\alpha - 2\pi) \times
({\rm area\,of\,horizon})\right) = {1\over 2}M\beta - {1\over 4} A,$$
where $A = 16\pi M^2$ is the area of the horizon. The total action is the
sum of the horizon and the asymptotic contributions,\footnote{Usually one
considers $M\beta = MT$ the action that an object of mass $M$ would have
if it had no entropy, so that the horizon contribution is only the
constant part, $A/4$.}
$$ I = M\beta - {A \over 4}. $$
Now $F = I/\beta = M - A/4\beta$ and $S = \beta^2\partial F/\partial\beta=
A/4$. So we get the same value as before, but with a better understanding
why the horizon area enters (it is just the extent of the horizon in the
two directions {\em not} shown in Fig.\ 14).

\bigskip
\unitlength .150mm
\begin{picture}(365.83,136.67)(-200,0)
\thicklines
\bezier{50}(351.67,90.83)(0.83,94.17)(0.00,0.00)
\bezier{50}(351.67,-90.83)(0.83,-94.17)(0.00,0.00)
\bezier{15}(340.83,45.00)(351.67,136.67)(362.50,45.00)
\bezier{10}(362.50,45.00)(365.83,0.00)(362.50,-45.00)
\bezier{15}(362.50,-45.00)(353.33,-135.00)(340.83,-45.00)
\bezier{10}(340.83,-45.00)(335.83,1.67)(340.83,45.00)
\put(0.00,0.00){\line(1,0){338.33}}
\bezier{250}(0.00,1.0)(0.00,-30.83)(340.00,-35.83)
\thinlines
\put(-24.17,-35.83){\vector(2,1){55.00}}
\put(-30.00,-36.67){\makebox(0,0)[rc]{$d\alpha$}}
\put(300,-17.50){\makebox(0,0)[lc]{$d\beta$}}
\put(339.5,-20.00){\vector(0,1){20.00}}
\put(339.5,-20.00){\vector(0,-1){14.17}}
\end{picture}
\vskip .7in

\noindent
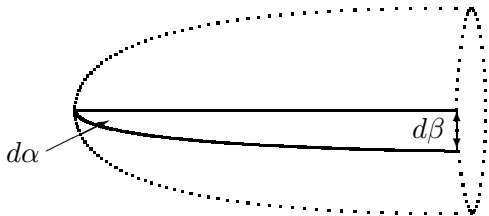
{\small {\bf Fig.\ 14.} A change in period by $dT=d\beta$ changes the action
by the amount enclosed in the heavy lines. Because these are $T$ = const,
$K=0$ surfaces, the change in the action is due only to the boundary $d\beta$
at infinity, and to the sharp corner of angle $d\alpha$ at the horizon.}

\bigskip \bigskip

We note that a similar picture allows us to interpret the black hole
entropy as entanglement entropy: if we propagate only half way around,
to $\beta = 4\pi M$, then the beginning and ending surfaces fit
together smoothly and have the geometry of a $T=0$ surface in a
(Lorentzian) Schwarzschild wormhole. The Euclidean surface enclosed
has no other boundary, it is therefore the WKB approximation to the
Hartle-Hawking ``no boundary'' state of the wormhole. Putting two
such ``half way around" surfaces together to the full surface of Fig.\ 11
amounts to (1) tracing over $\beta = 4\pi M$, i.e. the unobserved
``interior'' part of the wormhole (2) tracing over $\beta = 0$ to
get the entangled partition function from the propagator. Thus the entanglement
entropy is the same as what we computed above.

In this derivation an important role was played by the action due to
a sharp corner or ``joint'' of the boundary. Such corners are common
in Lorentzian regions of integration for the action, typically when
spacelike and timelike surfaces meet at right angles. In that case
one usually assumes that there is no contribution from such joints.
That this assumption is correct (but not for other angles, i.e.\ if
the scalar product of the normals to the spacelike and timelike
surfaces that meet does not vanish) has been shown by Hayward \cite{HA}.
But in Euclidean spaces one must associate a finite action with joints,
because these can be approached as limits of smooth surfaces. One
must also introduce special rules about joints in adjacent regions
so that the action is properly additive \cite{BH}: for example, two
$90^o$ corners, each with a positive value of $\int K$, can be put
together to form a ``$180^o$ corner,'' i.e. no corner at all, and
hence zero action contribution. In a sense the horizon area term in the
action is a result of insisting on proper additivity of Euclidean
actions with corners.

\bigskip \bigskip
\noindent
{\bf 6.2 Tunneling Interpretation}

\bigskip \noindent
Quantum mechanics allows a particle moving in a potential to tunnel into
classically forbidden regions. The WKB wavefunction for such a particle
can be obtained from the solution to a problem in classical physics,
namely the motion of that same particle in imaginary time. Since the
velocity is then imaginary, the kinetic energy is negative.\footnote
{If the total energy vanishes, the classical motion is the same as if the
particle had the usual, positive kinetic energy, and the potential were turned
upside down.} The transition between real- and imaginary-time motion occurs
at a classical turning point, where the particle's momentum vanishes.
At such a turning point the particle typically is most likely to be
reflected, as predicted by classical theory, but there is a smaller
probability that it makes a transition to imaginary-time motion until it
finds another turning point, where it emerges as a classical particle in
real time. Energy can be considered to be conserved during this whole
motion. The tunneling probability contains an exponential factor, which can be
computed from the action of the imaginary time (Euclidean) motion, and
a prefactor, which can be computed from the motion of nearby particles.
Although the Euclidean motion appears as a calculational device in quantum
mechanics, it does trace out the ``most probable escape path.''\footnote{One
can measure the particle's position along this path as long as
the time remains ill-defined so that no large amount of energy is
transferred to the particle.}

By analogy, Euclidean solutions of the Einstein equations represent such
tunneling histories of geometries. In the metric representation of the (as
yet unknown) quantum gravity, the wave functional $\Psi$ depends on spacelike
3-geometries. A nearly classical $\Psi$ is highly peaked about 3-geometries
that all fit as embedded subspaces into {\em one} 4D spacetime. Conversely,
a classical solution of the Einstein equations can be thought of as a
representation of $\Psi$: You consider the 3-geometries of all the possible
spacelike surfaces in the spacetime. $\Psi$ is large on those 3-geometries,
and vanishingly small on all others, that do not occur as subspaces of
the given spacetime. A Euclidean 4-space can be considered in a similar way
as a description of a $\Psi$ that is large on all 3-geometries that fit
into that Euclidean 4-space, and small on all others. Such wave functionals
describe classically forbidden transitions, for example topology change.
As for the particle, the beginning and end of the Euclidean motion occur at
initial data of vanishing momentum, i.e.\ of vanishing extrinsic curvature
(time-symmetric surface in the connecting Lorentzian spacetime).
Such 3-geometries are the only ones that can satisfy both the
Lorentzian and the Euclidean constraints, and are therefore suitable
surfaces for signature change. Euclidean geometries that describe a
transition to a real, Lorentzian spacetime (rather than just a virtual
fluctuation) must have such a surface of symmetry, usually called the
``bounce.''\footnote{Other Euclidean solutions are called ``instantons,''
but this term is often also applied to bounce solutions.}

Many instantons correspond to a classically stable initial state, for
example the vacuum, a classical background, or ``nothing.'' For a particle
at the stable position $x=0$ in a potential $V \sim x^2$, the Euclidean
motion is an infinitely slow roll-off from the upside-down potential
$V \sim -x^2$. Similarly, in an instanton geometry the initial state
is found in the asymptotic region. To represent a transition the
instanton must therefore have (1) the appropriate asymptotics (2) finite
action I, so that the WKB factor $e^{-I}$ in the transition amplitude
is finite (3) a bounce surface of symmetry and (4) a finite
prefactor. The prefactor is not easy to evaluate, so one usually is
satisfied to show that it is finite. The essential property is that
among the first order perturbations of the instanton there should be a
``negative mode,'' i.e. an eigenfunction of the action's second order
perturbation with negative eigenvalue. Thus one shows that there is a
zero mode that has a node, which indicates that there is a lower, hence
negative, eigenvalue.

The Euclidean Schwarzschild geometry is a typical example of a bounce
solution. To see its properties, rotate Fig.\ 11 about an axis at
$x = -2M$ as explained earlier, and look at a top view of the result,
down the $\rho$-axis (Fig.\ 15). Each point in the plane at $r > 2M$
now represents a circle in the $T$-direction. At $r=2M$ the circle
degenerates to a point, and at $r < 2M$ there is a ``hole,'' no points
in the Euclidean Schwarzschild geometry correspond to it. A possible bounce
surface of symmetry is the equatorial plane; this is the final state of the
tunnelling and the initial state of product of the transition. The
geometry on this surface is given by the metric of Section 2 with
$\theta = 0$. The initial state is in the asymptotically flat region,
so it is a flat cylinder of topology $\rm I\!\!R^2 \times S^1$.

\vskip 1in

\unitlength .500mm
\begin{picture}(120.00,64.00)(-70,0)
\put(70.00,50.00){\circle{12.00}}
\put(30.00,50.00){\line(1,0){34.00}}
\put(76.00,50.00){\line(1,0){34.00}}
\put(30.00,3.00){\line(1,0){80.00}}
\put(119.00,13.00){\vector(-2,-1){19.00}}
\put(120.00,60.00){\makebox(0,0)[lc]{bounce}}
\put(119.00,60.00){\vector(-2,-1){19.00}}
\put(120.00,13.00){\makebox(0,0)[lc]{initial state}}
\put(58.00,63.00){\vector(1,-1){8.40}}
\put(58.00,64.00){\makebox(0,0)[cb]{$r=2m$}}
\put(78.00,61.00){\vector(-2,-3){7}}
\put(79.00,61.00){\makebox(0,0)[lc]{hole}}
\thicklines
\bezier{20}(30,44)(70,44)(110,44)
\end{picture}

\bigskip
\noindent
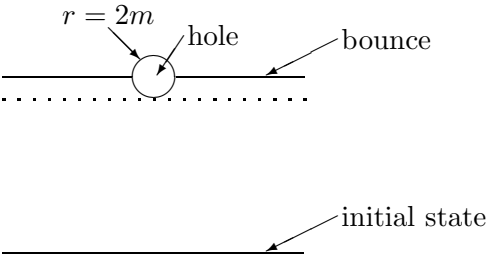
{\small {\bf Fig.\ 15.} The Euclidean Schwarzschild solution as a bounce,
starting in the asymptotically flat region, developing to the surface
(horizontal dotted line) where the hole first appears, and to the
bounce surface where the curvature and the hole radius reach their maximum;
followed by the symmetrical development back to flat space.}

\bigskip \bigskip
What physics does this bounce instanton describe? Its action
(the Euclidean Schwarzschild
action) is $< \infty$, as we saw. There is a zero mode --- the active
coordinate transformation of the metric that shifts everything in Fig. 15
rigidly upward. An infinitesimal shift has a node at the symmetry (bounce)
plane. Hence there should be a negative mode, and the instanton describes
a quantum transition of non-vanishing probability between the initial and
the final state. The initial state is flat space, which is classically
stable. It is somewhat strange because it is compact in one direction
(the $T$-direction which, in spite of its name, is spacelike); but if
we repeat the model one dimension higher, based on the 5-dimensional
Schwarzschild solution, the initial state of type $\rm I\!\!R^3 \times S^1$
can be interpreted as a Kaluza-Klein vacuum, compactified to length
$\sim M$ in the extra direction. In either dimensionality the corresponding
instanton connects the initial classical vacuum to a decay product, which
starts its Lorentzian time development at the bounce surface with zero
extrinsic curvature. The topology of the bounce surface is
$\rm I\!\!R^2 \times S^1$ for the 4D bounce, and $\rm I\!\!R^2 \times S^2$
in the 5D case, so in the latter case a topology change is certainly
involved. Its Lorentzian time development is another analytic continuation
of the instanton metric, via $\theta \rightarrow i\theta$. The circular
Euclidean hole becomes a hyperbolic timelike hole in the continuation,
a hole that expands at constant acceleration in all directions. Thus
the compactified vacuum is unstable to decay into an expanding hole,
with the initial hole radius and its acceleration determined by $M$, i.e.\
by the compactification scale \cite{W}. Note that the actual decay history
corresponds to only half of the bounce geometry, up to the bounce surface.
Beyond that surface the signature changes and the Lorentzian development
takes over.

Some instantons, when cut in half to reveal the bounce surface, have no
other boundary or infinite region. These satisfy the Hartle-Hawking
``no boundary'' condition. In this case there is no initial state, which
has been called ``creation from nothing." The simplest example is the
$S^4$ Euclidean de Sitter solution, the round 4-sphere. An equatorial
3-space is the bounce moment, and the hemisphere which bounds it has
no other boundary. In this view one can imagine the early history of the
universe as in Fig.\ 15, a Euclidean ``time-less'' development of
a bounce surface, where the signature changes to Lorentzian and de Sitter
inflation starts.

\bigskip
\unitlength .500mm
\begin{picture}(128.94,78.73)(-70,0)
\thicklines
\bezier{30}(90.00,30.00)(89.67,18.44)(80.00,12.78)
\bezier{30}(80.00,12.78)(70.00,7.22)(60.00,12.78)
\bezier{30}(60.00,12.78)(50.33,18.44)(50.00,30.00)
\bezier{100}(90.00,30.33)(91.33,48.00)(113.00,69.67)
\bezier{100}(50.00,30.33)(48.67,48.00)(27.00,69.67)
\bezier{30}(57.67,28.00)(42.33,30.67)(57.67,32.00)
\bezier{20}(57.67,32.00)(68.67,34.00)(82.33,32.00)
\bezier{30}(82.33,32.00)(97.00,30.00)(82.33,28.00)
\bezier{20}(82.33,28.00)(70.00,26.00)(57.67,28.00)
\bezier{60}(43.08,65.63)(9.61,71.46)(43.08,75.5)
\bezier{50}(43.08,75.5)(67.09,78.73)(96.92,74.37)
\bezier{60}(96.92,74.37)(128.94,70.00)(96.92,65.63)
\bezier{50}(96.92,65.63)(70.00,61.27)(43.08,65.63)
\put(120.00,20.00){\vector(0,1){10.00}}
\put(120.00,20.00){\vector(0,-1){10.00}}
\put(122.00,20.00){\makebox(0,0)[lc]{Euclidean}}
\put(120.00,50.00){\vector(0,1){20.00}}
\put(120.00,50.00){\vector(0,-1){20.00}}
\put(122.00,50.00){\makebox(0,0)[lc]{Lorentzian}}
\put(38,30){\vector(1,0){12}}
\put(36,30){\makebox(0,0)[rc]{bounce surface}}
\end{picture}

\noindent
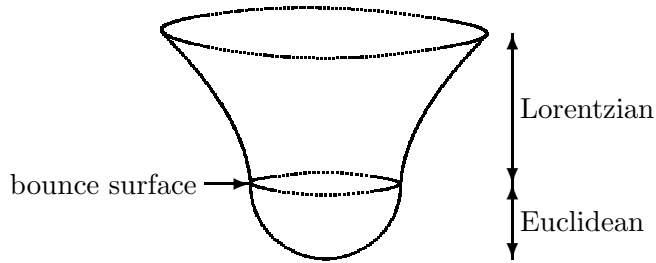
{\small {\bf Fig.\ 16.} The de Sitter universe created as a quantum event
from nothing. It is tempting to think of the lowest point as the ``nothing''
from which the universe starts, but this has no invariant meaning, since
the whole figure is invariant under the rotation/de Sitter group.}

\bigskip \bigskip
There are other universes that can be created from nothing if there is
a cosmological constant, for example the Nariai universe. It is similar
to the  wormhole de Sitter universe (Fig.\ 7 with the two throats at $R_o$
identified), except that it has the same size throughout (limit $R_o
\rightarrow R_c$).\footnote{This is similar to the limit discussed at the end
of Section 3.} However, the Euclidean action of these other universes is
larger than that of de Sitter. They therefore contribute to the
Hartle-Hawking state with a lower probability amplitude.

A very interesting set of instantons are those connecting a background field
configuration with a bounce surface that contains a pair of black holes
\cite {GS}. The background is a magnetic (or electric) field, and the
instanton describes pair production of magnetically (or electrically)
charged black holes by this field. This instanton metric is rather complicated,
but a similar process occurs in --- again --- the de Sitter universe. Here the
cosmological constant takes the place of the background field that pulls
the virtual particles apart sufficiently to make them real. In fact, the
Nariai solution mentioned above can be thought of in these terms, and the
RNdS solutions provide other examples (where we must set either set $Q^2=M^2$
to
avoid Euclidean singularities, or use extremal black holes whose periodicity
is arbitrary and can therefore be set equal to that of the cosmological
horizon). Fig.\ 13 shows half of such a $Q^2=M^2$ instanton,
up to the bounce surface. It is a creation from nothing because there
is no other boundary than the bounce surface, but in this case it is
clear that there is no preferred point of ``nothing'' from which to start
the universe. Note that not only the wormhole geometry but also its
electric or magnetic field lines are created from nothing. The Euclidean
action of these instantons has been calculated, so one can
estimate the probability of the various outcomes (Nariai-type vs $Q^2=M^2$
type vs extremal type, different $M$-values) relative to plain de Sitter
space \cite {MR}. Generally the smaller the wormhole, the greater its
probability.

There are also instantons that start from something rather than nothing,
or vacuum, or background. These describe the type of quantum fluctuations often
envisaged by Wheeler, whereby one extremally charged wormhole in
asymptotically flat space breaks up into several (or vice versa).
The instanton \cite{DB} is analytically related to the Majumdar-Papapetrou
solution for any number of extremally charged black holes, arbitrarily
placed. This metric and field have the form
$$ds^2 = -V^{-2}dt^2 + V^2 d\sigma^2 \qquad A = V^{-1}dt$$
where $d\sigma^2$ is the metric of flat 3-space, and $V$ is a solution
of the Laplace equation in this 3-space, $\nabla^2 V = 0$. To obtain the
solution for several black holes in asymptotically flat space,
one puts $V = 1 + \sum M_i/r_i$. But here we consider instead the Euclidean
solution obtained by $t \rightarrow it$ and choosing a somewhat different $V$,
\begin{equation}
ds^2 = V^{-2}dt^2 + V^2 d\sigma^2 \qquad V = \sum M_i/r_i
\end{equation}
In the limit $r \rightarrow \infty$ this describes a cylindrical space
of size $\sum M_i$, whereas in each limit $r_i \rightarrow 0$ we get
a cylindrical space of size $M_i$. The instanton interpolates between
these, as shown in Fig.\ 17.

\bigskip \bigskip
\unitlength 1.00mm
\begin{picture}(114.00,50.00)
\thicklines
\put(65.00,30.00){\line(0,-1){25.00}}
\put(88.00,30.00){\line(0,-1){25.00}}
\bezier{84}(70.00,27.00)(60.00,30.00)(70.00,33.00)
\bezier{84}(83.00,33.00)(93.00,30.00)(83.00,27.00)
\bezier{52}(70.00,33.00)(76.50,34.33)(83.00,33.00)
\bezier{52}(70.00,27.00)(76.50,25.67)(83.00,27.00)
\bezier{28}(70.00,30.00)(72.83,32.33)(75.00,30.00)
\bezier{24}(70.00,29.50)(72.50,27.50)(75.00,29.33)
\bezier{28}(83.00,30.00)(80.17,32.33)(78.00,30.00)
\bezier{24}(83.00,29.50)(80.50,27.50)(78.00,29.33)
\thinlines
\bezier{28}(70.00,5.00)(72.83,2.67)(75.00,5.00)
\bezier{10}(70.00,5.50)(72.50,7.50)(75.00,5.67)
\bezier{28}(83.00,5.00)(80.17,2.67)(78.00,5.00)
\bezier{10}(83.00,5.50)(80.50,7.50)(78.00,5.67)
\thicklines
\bezier{30}(70.00,8.00)(60.00,5.00)(70.00,2.00)
\bezier{30}(83.00,2.00)(93.00,5.00)(83.00,8.00)
\bezier{52}(70.00,2.00)(76.50,0.67)(83.00,2.00)
\bezier{4}(75.00,8.17)(76.50,8.50)(78.00,8.17)
\bezier{30}(88.00,30.00)(87.00,46.00)(114.00,50.00)
\bezier{30}(65.00,30.00)(66.00,46.00)(39.00,50.00)
\thinlines
\put(56.00,19.00){\vector(2,1){9.00}}
\put(55.00,19.00){\makebox(0,0)[rc]{initial}}
\put(76.50,35.00){\vector(-1,-4){1.46}}
\put(76.50,35.00){\vector(1,-4){1.46}}
\put(76.50,35.67){\makebox(0,0)[cb]{final}}
\put(78.00,26.00){\line(0,-1){21.00}}
\put(70.00,27.00){\line(0,-1){22.00}}
\put(75.00,26.00){\line(0,-1){21.00}}
\put(83.00,27.00){\line(0,-1){22.00}}
\thicklines
\put(70.00,30.00){\line(0,-1){3.00}}
\put(75.00,30.00){\line(0,-1){4.00}}
\put(83.00,30.00){\line(0,-1){3.00}}
\put(78.00,30.00){\line(0,-1){4.00}}
\end{picture}

\bigskip
\noindent
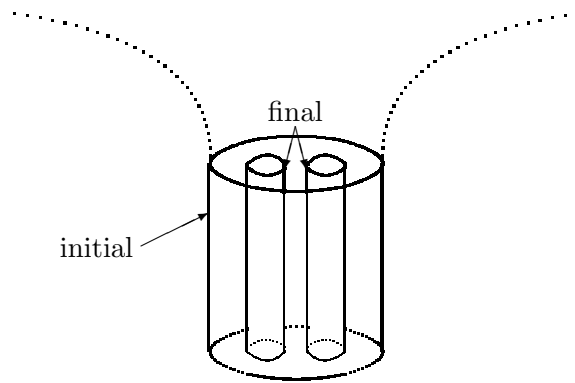
{\small {\bf Fig.\ 17.} Splitting of a Bertotti-Robinson universe into
two, as an approximation of the splitting of an extremal Reissner-Nordstr\"om
black hole. The dotted curve indicates the original black hole geometry,
which approaches the Bertotti-Robinson geometry far down the throat.
The $t$-direction of Eq (10) is plotted vertically. One angular direction
is suppressed.}

\bigskip \bigskip

This is not a bounce instanton, because it is not cut in half by a bounce
surface; it approaches the initial {\em and} final state asymptotically.
It can therefore be regarded as describing a tunneling fluctuation, similar
to the ground state in a double-well potential. The initial and final
cylindrical geometries, when continued to Lorentzian spacetimes, are
Bertotti-Robinson (BR) universes. Thus this instanton describes the splitting
of such universes, but because the geometry deep down in the throat
of extremal black holes approaches the BR geometry, it
can also be regarded as a description of the splitting of an extremal
black hole's throat.

The action of the instanton of Fig.\ 17 turns out to be entirely in the
joints that the initial and final surfaces make with the horizontal
surfaces. The interior does not contribute because the gravitational
and electromagnetic contributions cancel (as is reasonable in a $Q^2=M^2$
situation, where gravity is balanced by electromagnetism). The surfaces
outside the joints have $K=0$. The contribution of the jth joint is
$A_j/16$, where $A_j$ is the corresponding (horizon) area, and there
are two joints for each horizon. Thus the total action is
$$I = {\pi\over 2}\left[\left(\sum M_j\right)^2 - \sum M_j^2\right].$$
This action is independent of the axial length of the BR universes, or
of the depth of the throats. It is also curious that $I$ is half the difference
of the entropies of the corresponding black holes (at least according to
one prescription for calculating this entropy). Thus the square of the
WKB wavefunction $e^{-2I}$ agrees with the probability $e^{\Delta S}$
asssociated in statistical mechanics with a fluctuation in which the
entropy $S$ deviates by $\Delta S$ from its equilibrium value.

\bigskip \bigskip \noindent
{\large\bf 7 The Multi-Black-Hole Solutions}

\bigskip \noindent
Can analytic continuation from regular initial data lead to naked
singularities? Cosmic Censorship forbids it, so one simple way to
settle the question would be to find a counterexample. Consider, for
example, the RNdS universe of Figs.\ 6 and 7.
In this universe (as in the $\Lambda=0$, asymptotically flat
Reissner-Nordstr\"om (RN) geometry) a geodesic observer that
wants to experience the singularity can do so, for example by moving
along the vertical symmetry axis of the figure. In the RN case this
is not considered a serious challenge to cosmic censorship, because
the interior of the black hole, through which the observer in search
of a singular experience must travel, is not stable under small perturbations
of the exterior: radiation falling into the hole from the exterior
would have a large blueshift at this observer --- it would not only
burn her up, but also change the nature of the singularity. It is
remarkable that this does not necessarily happen in RNdS
universes, for certain values of the parameters \cite{MM}.
Thus these solutions
are a counterexample to a strong interpretation of cosmic censorship.
But there are, of course, many other geodesics that can lead observers
who do not take the plunge to their safe haven at $\Scri^+$.

For other parameters (the overmassive case) RNdS does have naked
singularities: an observer at $\Scri^+$ in Fig.\ 9 sees them. But these
singularities exist for all times, they do not arise from regular initial
data. Can we find spacetimes with black holes
that change from undermassive to overmassive? Progress on this
question has been made possible by solutions of the Einstein-Maxwell
equations that can describe merger of black holes.

A solution representing any number $n$ of arbitrarily placed charged black
holes in a de Sitter background is given by a metric of type (6), with
a different potential $U$ \cite{KT}:
\begin{equation}
ds^2=-{d\tau^2\over U^2}+U^2(dr^2+r^2d\Omega^2) \qquad
U = H\tau+\sum_{i=1}^n{M_i\over |r-r_i|} \qquad A_\tau={1\over U}\,.
\end{equation}
We will call this the KT solution.  Each mass of the KT solution has
a charge proportional to the mass, $Q_i=M_i$, and only
the location $r_i$ (not the initial velocity) is arbitrary. Here
$|r-r_i|$ denotes the Euclidean distance between the field point $r$
and the fixed location $r_i$ in a Euclidean space of
cosmological coordinates. For $n=1$
this reduces to the $Q^2=M^2$ case of the RNdS solution (6). Also, in the
limit of large $r$, and for $r_i$ in a compact region of Euclidean
coordinate space, (11) approaches the RNdS solution with $M = \sum M_i$.
One therefore expects the horizon structure at $r \rightarrow \infty$ to
be similar to that of RNdS, which suggests that a $H>0$ and a $H<0$ version
of (11) can be glued together as extensions of each other, similar to
Fig.\ 8. The surprise is that, although this can be done with some degree
of smoothness, it cannot be done analytically \cite{BHKT}.
This means that there
is no unique extension. An observer at rest in contracting ($H<0$)
cosmological coordinates, whose entire past can be described in these
coordinates, has no way of telling what is in the other ``half'' of
the de Sitter background  (he can only guess that the total charge over
there must be $-\sum M_i$, to balance the charge that he sees). If he moves
and crosses the cosmological horizon $r=\infty$, the other, expanding half
suddenly comes into view, seen at a very early time when all the masses are
very close together. It is therefore reasonable
that there is a pulse of gravitational and EM radiation associated with
the horizon: this is the physical description of the lack of analyticity.

It is instructive to note how the various solutions differ in respect to
possible coordinate choices. Pure de Sitter space has a static and two
cosmological (expanding and contracting) coordinate systems centered
about any timelike geodesic. In
RNdS these coordinates are centered about the black
hole only, but one can still choose between expanding and contracting
frames near either of the holes. In KT there is one set of black holes,
all with the same sign of charge, that is uniquely expanding (the distances
between holes increasing as $e^{H\tau}$), and another, oppositely
charged set that is uniquely contracting (the distances decreasing as
$e^{-|H|t}$). The expanding set is described by cosmological coordinates
that include $\Scri^+$, whereas for the contracting set, $\Scri^-$ is
included.

\bigskip \bigskip \noindent
{\bf 7.1 Merging Black Holes}

\bigskip \noindent
Since black (as opposed to white) holes are determined from $\Scri^+$, it is
easiest to identify the black hole horizons for the expanding set. Consider
the expanding coordinates in RNdS as in the left half of Fig. 7. The
boundary of the past of $\Scri^+$ that lies in those coordinates is the
left-hand line labeled $r=0$ --- it is the event horizon of the black hole
in the ``left'' part of RNdS space. Similarly the event horizon of
expanding KT space is given by $r - r_i = 0$. The Euclidean coordinate
space with the $n$ points $r_i$ removed is a representation of $\Scri^+$
of expanding KT space. The $n$ missing points represent $n$ disjoint
boundary components, and there is a finite distance between them at all
times. Thus we have $n$ black holes that remain separate for all times.

It is not as easy to identify the black holes in the contracting KT space,
because that does not contain much of $\Scri^+$. But because the cosmological
horizon at $r=\infty$ is so similar to that of the RNdS space it does contain
a ``point'' like S of Fig.\ 8 that is just enough to define the event
horizon. So, to find the event horizon while staying within one coordinate
patch we must find the surface that divides lightrays reaching $r=\infty$
from those that reach the singularity. But where in the KT world
is the singularity? One can show that
metrics of type (11) are singular where $U=0$. In the contracting case,
$H<0$ (and of course $M_i>0$), this happens only for positive $\tau$.
Thus we need to find the ``last" lightrays that just make it out to
$r=\infty$ at $\tau=0$. More precisely (since $r=\infty$ is not a very
precise place) we need $H r\tau$ finite in this limit.\footnote{This
is so because part of the ``somewhat involved" coordinate transformation
leading to (5) is actually rather simple, $R=Hr\tau+M$. For RNdS this
is the static $R$, which is finite on the black hole horizon. Because
near $r=\infty$ the KT geometry is so close to RNdS, $R$ should also
be finite for KT.}

If we know these last lightrays (the 3D horizon surface) we can then
intersect them with a spacelike surface to find the shape of the 2D
horizon at different times. An interesting question is whether this
2D horizon changes topology with time. This would, for example, describe
the merger of two black holes into one. If black holes merge we can try
to make an overmassive (hence nakedly singular) one from two undermassive
ones, to test cosmic censorship. This is of course possible only in a
contracting part of the KT solution, because we have already seen that
the black holes in the expanding part remain separate for all times.

All contracting black holes do eventually merge into one. To show this
for the case of a pair of holes, we show that the horizon must consist
of two parts at early times, and be a single surface at late times. One
can see rather directly that light starting at $r$ sufficiently close to
any one of the $r_i$ at any time and in any direction will reach the
singularity, because it will spend its entire history in a geometry
sufficiently close to the single black hole, RNdS geometry. Thus points
sufficiently close to $r_i$ will always lie within the event horizon.
So, for the case of two black holes, a key question is what happens to
light that starts on the midplane between the holes. In fact, if the light
starts early enough it will always be able to escape to $r=\infty$. Thus
at early times the midplane does not meet the horizon --- the two black
holes are disjoint. To show this we center our Euclidean coordinates
at the midpoint between the holes, which have a Euclidean separation $d$.
{}From $ds^2=0$ in (11) we then find, for radial outgoing null geodesics in the
midplane,
$${d\tau\over dr} = U^2 = \left(H\tau+{M\over\sqrt{r^2+d^2}}\right)^2
< \left(H\tau+{\sqrt{2}M\over {r+d}}\right)^2.$$
Now let $$R_*=H\tau (r+d) \qquad y_*=\ln(r+d)$$ to find
\begin{equation}
(r+d)\, dR_*/dr > R_* +H\,(R_*+\sqrt{2}M)^2.
\end{equation}
Standard analysis of this equation shows that if $R_*$ is larger than
the lower root of the RHS of (12), it will stay positive
for all larger $r$. In terms of $r$ and $\tau$ this means that if $\tau$
is sufficiently negative for any $r$ (remember $H<0$) then $\tau$ will
remain negative as $r$ increases --- the lightray avoids the singularity.
By a similar estimate one can show \cite{BHKT}
that for each sufficiently late
(but negative) $\tau$ there is a sphere surrounding both $r_i$ such
that all outgoing null geodesics will reach the $U=0$ singularity. This
means that the horizon surrounds both $r_i$, and the black holes have
merged.

\bigskip \bigskip
\noindent
{\bf 7.2 Continuing Beyond the Horizons}

\bigskip \noindent
We could now discuss the merging of two undermassive into one overmassive
black hole. Since the geometry near $r=\infty$ will be determined by the
overmassive sum of the two individual masses, that neighborhood will
look like the corresponding part of a single overmassive RNdS, that is,
like the left hand side of Fig.\ 8. That contains a naked singularity
(the left multiply-crossed line), which has nothing to do with the black
hole merger, because it is located at the opposite side of the universe.
But this is the only place in the coordinate patch where a $U=0$ singularity
occurs. Our patch does not describe enough of the history of the interesting
region, where $|r-r_i|$ is small, just as the thick outline does not extend
far to the right side of Fig.\ 8. To find out whether black hole merger
generates its own naked singularity we must continue the KT metric beyond
the inner black hole horizons at $r=r_i$. As we are continuing
the KT geometry it is of course also interesting to look beyond the
cosmological horizon at $r=\infty$, which exists only if the total mass
is undermassive ($4|H|\sum M_i<1$). So we look at null geodesics that
approach these horizons.

Let us choose the origin $r=0$ of our Euclidean coordinates at the location
of the $i^{\rm th}$ mass (so that $r_i=0$). The equation $ds^2=0$ satisfied
by an ingoing null geodesic then takes the form, for small $r$,
\begin{equation}
 {d\tau\over dr}=-U^2=-\left(H\tau+{M\over r}+
\sum_{j\neq i}{M_j\over r_j}\right)^2.
\end{equation}
We can eliminate the last (constant) term on the right by defining a new
time coordinate $\tau'=\tau +H^{-1}\sum'(M_j/r_j)$. The equation then becomes
an equality version of (12), and by analyzing it in the same way as above one
finds \cite{BHKT} the limiting forms
\begin{equation}
2H^2r\tau' \rightarrow 1-2M_iH-\sqrt{1-4M_iH}.
\end{equation}
To assess any incompleteness we need to know how a null geodesic
$(r(s),\,\tau(s))$ depends on the affine parameter $s$, and we can get that
from the variational principle,
$$\delta \int \left(-{1\over U^2}\left({d\tau\over ds}\right)^2 +
U^2\left({dr\over ds}\right)^2\right)ds=0.$$
The Euler-Lagrange equation for $\tau(s)$ together with the first equality
of (13) yields
$${{d^2r}\over{ds^2}}-2HU\left({dr\over ds}\right)^2=0.$$
Substituting (14) we find, in the limit $r \rightarrow 0$,
$${d^2r\over ds^2}-{1-\sqrt{1-4M_iH}\over r}\left({dr\over ds}\right)^2=0$$
with the solution
\begin{equation}
r \sim \left(s-s_{\rm hor}\right)^{1\over\sqrt{1-4M_iH}}, \quad {\rm hence}
\quad \tau\sim \left(s-s_{\rm hor}\right)^{-{1\over\sqrt{1-4M_iH}}}.
\end{equation}
So the inner horizon is reached at a finite parameter value $s_{\rm hor}$.
Similarly one finds that the cosmological horizon is reached in a
finite parameter interval,
\begin{equation}
r \sim\left(s-s_{\rm Hor}\right)^{-{1\over\sqrt{1+4(\Sigma M_i)H}}}, \qquad
\tau \sim \left(s-s_{\rm Hor}\right)^{1\over\sqrt{1+4(\Sigma M_i)H}}.
\end{equation}

This behavior of the coordinates $r$ and $\tau$ gives us important information
about the differentiability and analyticity of the geometry near the horizon.
We can first eliminate whichever of the two is infinite on a given horizon
in favor of $\hat R=Hr\tau$, which is always finite on the horizon.
The metric is then an analytic function of remaining, finite
coordinates, so the Riemann tensor will also be analytic
in these coordinates. But in order that the geometry
be differentiable, the Riemann tensor should be differentiable
in the affine parameter $s$ along null geodesics. Thus the
differentiability of the geometry is measured by that of
$r$ resp.\ $\tau$ as a function of $s$, as given by (15) and (16).

Consider first the neighborhood of the inner horizon, where $r$ is finite.
Since $H<0$ we have $1/\sqrt{1-4M_iH} < 1$, $r$ is not a differentiable
function of $s$ at $r=0$, and the Riemann tensor will be singular there.
A more careful analysis \cite{BHKT}, using a transformation to coordinates
that are not singular on the horizon, shows that the metric is $C^1$ but
in general not $C^2$. There is therefore no unique, analytic extension across
the inner horizon. One can match differentiably essentially any KT solution
with the same mass $M_i$.
One can increase the differentiability by arranging
the other masses carefully around the $i^{\rm th}$ one, so as to make
the potential $U$ approximately spherically symmetric (by eliminating
multipoles to some order). The neighborhood of $M_i$ then becomes
approximately RNdS and hence ``more nearly analytic'' --- i.e., of
increased differentiability.

The situation near the cosmological horizon offers more variety. To have this
horizon at all the total mass must be undermassive, $4|H|\Sigma(M_i)<1$.
Here $\tau$ is the finite one, and the corresponding power of $s$ is
$1/\sqrt{1-4(\Sigma M_i)|H|}>1$. Thus $\tau$ is always at least $C^1$.
The transformation
to coordinates that are good on the horizon shows that the metric is
always at least $C^2$. In the special
cases when the power is an integer $n$, i.e., for masses such that
$$4H\sum M_i = 1 - {1\over n^2},$$
the metric is $C^\infty$. For these values the smooth continuation matches
the KT spacetime at the cosmological horizon to one with the same
position and magnitudes of all the masses (so that all multipole moments
agree), but with the opposite sign of $H$. We do not understand
the physical significance of these special masses.

To show that the geometry at these horizons is not more differentiable
than claimed one can compute the Riemann tensor. This infinity is of the
null type mentioned in Sect.\ 2, and does not show up in invariants
formed from the Riemann tensor. To see this for the horizon at $r=0$
(or $r=\infty$) we write the metric (11) in terms of the coordinate
$\hat R=Hr\tau$ and $y=\ln r$, and the quantity $W=rU$. Then the horizon occurs
at $r \rightarrow \pm \infty$, where $\hat R$ and $W$ are finite,
$$ds^2 = -{(d\hat R-\hat Rdy)^2\over H^2W^2}+W^2(dy^2+d\Omega^2).$$
Now an invariant formed from the curvature tensor involves terms
in derivatives of the metric and its inverse, multiplied by powers
of the metric and its inverse. All these reduce to derivatives of $W$
and $\hat R$ divided by powers of $W$. But all derivatives of $W$ remain
bounded
as $y\rightarrow \pm\infty$, and $W$ is finite on the horizon. Thus the
invariants cannot blow up.

Singularities do show up in the components of the Riemann tensor in a
parallelly propagated frame, for example along the null geodesic
$(r(s),\,\tau(s))$ discussed above. Let $l=\partial/\partial s$ be
the parallelly propagated tangent. Because of the asymptotic symmetry
near the horizon, $\eta = \partial/\partial\theta$ is also asymptotically
parallelly propagated. Now the frame component
$R_{\mu\nu\rho\sigma}l^\mu\eta^\nu l^\rho\eta^\sigma$ contains the term
$g_{\theta\theta,ss}$. If $g_{\theta\theta}=W^2$ depends on $r$ (and not just
on the regular $\hat R$), it will not be a smooth function of $s$. In the
RNdS (``single mass'') case, $g_{\theta\theta}$ depends only on $\hat R$.
In the KT case the corrections to that behavior near an inner ($r=0$) horizon
start with the power $r^2=\left(s-s_{\rm hor}\right)^{2\over\sqrt{1-4M_iH}}$
unless there is special symmetry; thus one finds the differentiability
of the metric as claimed above.

\bigskip\bigskip

{\large\bf\noindent
8 Naked Singularities?}

\bigskip \noindent
Naked singularities visible to observers safely outside the strong
curvature regions do not form in realistic gravitational collapse ---
this is the essential notion behind cosmic censorship. It can be made more
precise in various ways \cite{xx}. At present none of these have been proved
to be true in generic cases. When a proof seems difficult, it may be
easier to obtain a convincing counterexample. Even if the conjecture is correct
under certain assumptions, counterexamples are useful
to test the necessity of these assumptions. For example, it may be that a
version of cosmic censorship holds in pure general relativity, but
fails when the theory is modified, say by a cosmological constant or
by applying it to higher dimensions. There are in fact indications that
cosmic censorship fails in the higher-dimensional theory inspired by
string theory:  certain 5D black strings (objects that appear
four-dimensionally like black holes) are unstable, it is entropically
favorable for them to decay into a set of 5D black holes, and
during this decay naked singularities would form \cite{GF}.
In the present contribution we want to test whether cosmic censorship
fails in the special circumstances that are afforded when a cosmological
constant is present.\footnote{A similar test for Einstein-Maxwell-dilaton
theory with a cosmological constant inspired by string theory has been
discussed in \cite{HH}.}

The idea of using KT spacetimes to test cosmic censorship is to start
with two or more small (and hence not naked) black holes and
let them collapse to form a single large (and maybe naked) one.
We have seen that the KT solutions indeed can describe coalescing
black holes, in the sense that the event horizons coalesce. But if
it is possible to define the event horizon as we did, by observers who
live for an infinite proper time in one KT coordinate patch, then these
observers will see no signal from any singularity
--- all singularities that form from
the initial data, including those in any of the (non-analytic) extensions,
lie inside this event horizon. To find a situation where there is no
``safe haven,'' so that the generic observers does see a singularity,
we must suppose that $\sum M_i$ is overmassive, so the initial black
holes cannot be defined by their event horizon. An alternative to
starting with black holes that have event horizons is to start with regular
initial data on a compact surface. The KT solution cannot provide
this either, because each $M_i$ has an infinite throat. But such throats
are the next best thing: each throat is undermassive, it is surrounded
by a trapped surface,
so one would not expect that the asymptotic regions down the throat
could influence the solution in the interior. Can we, then, construct
a KT solution of undermassive throats that has regular initial data
and a naked singularity in its time development?

To decide this we must explore the global structure of the KT geometry.
Unfortunately this cannot be completely represented by 2D conformal
diagrams, because there is insufficient symmetry to suppress the
additional dimensions. If we confine attention to the case of two
equal masses, they lie on a line in the Euclidean coordinate
space (which is an axis of symmetry of the spacetime). We can represent
the essential features of the spacetime by drawing the conformal
diagram for the spacetime spanned by the part of the axis going from
one of the masses to infinity. This is shown in Fig.\ 18a.
The part of the diagram
representing the region near $r = \infty$ is to be read like
a normal conformal diagram (i.e., each point represents a 2-sphere),
whereas the region near $r=0$ is to be thought of as doubled (each
point represents two 2-spheres).

\unitlength=1.30mm
\begin{picture}(50.00,53.50)(20,4)
\thinlines
\put(10,0){
\multiput(0,0)(0,1.2){25}{
\put(20.00,10.00){\makebox(0,0)[cc]{{\bf*}}}}
\multiput(20,10)(0,1.2){25}{
\put(20.00,10.00){\makebox(0,0)[cc]{*}}}
\put(20.00,40.00){\line(1,1){10.00}}
\put(30.00,50.00){\line(1,0){10.00}}
\put(15.00,25.00){\vector(1,0){4.50}}
\put(14.00,25.00){\makebox(0,0)[rc]{$r'=0$}}
\put(30.00,20.00){\makebox(0,0)[rc]{$r=0$}}
\put(31.00,21.00){\vector(1,1){4.00}}
\put(31.00,19.00){\vector(1,-1){4.00}}
\put(25.00,10.00){\makebox(0,0)[ct]{$\Scri^-$}}
\put(35.00,50.00){\makebox(0,0)[cb]{$\Scri^+$}}
\thicklines
\put(20.00,10.00){\line(1,0){10.00}}
\put(30.00,10.00){\line(1,1){10.00}}
\put(40.00,20.00){\line(-1,1){20.00}}
}
\thinlines
\put(80.00,10.00){\line(1,0){10.00}}
\put(90.00,10.00){\line(1,1){10.00}}
\put(85.00,10.00){\makebox(0,0)[ct]{$\Scri^-$}}
\put(82.00,24.00){\vector(1,0){4.00}}
\put(81.00,24.00){\makebox(0,0)[rc]{$r'=0$}}
\put(102.00,21.00){\makebox(0,0)[lc]{$r=0$}}
\put(102.00,20.00){\vector(-3,-2){6.00}}
\put(102.00,22.00){\vector(-3,1){6.00}}
\put(82.00,11.00){\circle*{0.00}}
\put(84.00,11.00){\circle*{0.00}}
\put(82.00,13.00){\circle*{0.00}}
\put(84.00,13.00){\circle*{0.00}}
\put(82.00,15.00){\circle*{0.00}}
\put(84.00,15.00){\circle*{0.00}}
\put(88.00,25.00){\circle*{0.00}}
\multiput(80,10)(0,1.2){17}{
\put(20.00,10.00){\makebox(0,0)[cc]{*}}}
\put(40.00,7.00){\makebox(0,0)[ct]{a}}
\put(91.00,7.00){\makebox(0,0)[ct]{b}}
\bezier{152}(80.00,10.00)(81.00,21.00)(100.00,40.00)
\bezier{140}(85.00,10.00)(87.00,24.00)(100.00,40.00)
\put(100.00,20.00){\line(-1,1){9.67}}
\thicklines
\bezier{84}(100.00,20.00)(90.00,14.00)(81.00,14.00)
\thinlines
\put(84.00,17.00){\circle*{0.00}}
\put(86.00,17.00){\circle*{0.00}}
\put(84.00,19.00){\circle*{0.00}}
\put(86.00,19.00){\circle*{0.00}}
\put(85.00,21.00){\circle*{0.00}}
\put(87.00,21.00){\circle*{0.00}}
\put(86.00,23.00){\circle*{0.00}}
\put(88.00,23.00){\circle*{0.00}}
\put(89.00,27.00){\circle*{0.00}}
\put(91.00,28.00){\circle*{0.00}}
\put(92.00,30.00){\circle*{0.00}}
\put(93.00,32.00){\circle*{0.00}}
\put(94.00,33.00){\circle*{0.00}}
\put(95.00,34.00){\circle*{0.00}}
\put(90.00,26.00){\circle*{0.00}}
\end{picture}

{\noindent \small
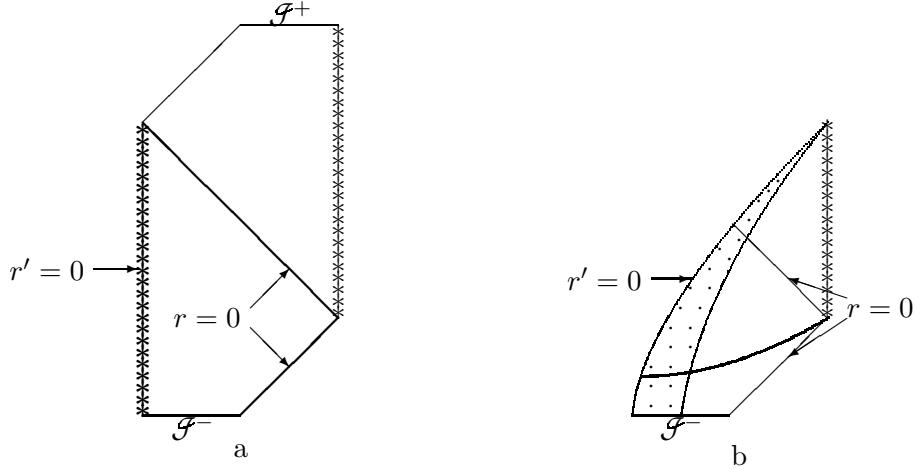
{\bf Fig.\ 18a.} Conformal diagram of history of axis from one black hole
to ``infinity'' for a two-black-hole contracting KT geometry. The black
holes correspond to the region on the right, and the surrounding ``de
Sitter background'' is the region on the left. The total mass is assumed
to be overmassive, so this ``background'' is really an overmassive RNdS
geometry with the singularity shown on the left. Therefore the axis
extends to $r=\infty$ only for $\tau <0$, after that it hits this singularity
at $U(r,\tau)=0$. The $C^1$ extension across the upper $r=0$ horizon
was chosen to be the time-reverse of the heavily outlined region.

{\bf Fig.\ 18b.} In this diagram the cosmological singularity of Fig.\ 18a
is covered up by a sphere of dust, as discussed in Sect.\ 4.1. The dust
region is shown dotted. The thick curve is a spacelike surface with
nonsingular initial data, containing two infinite charged black hole
throats and, on the antipodal point of the universe, a collapsing sphere
of dust. All observers have to cross the future $r=0$ Cauchy horizon and
can thereafter see the naked singularity shown on the right, the result
of the merger of the two black holes. The spacetime ends after a finite
proper time in a ``big crunch.''
}

\bigskip\bigskip

The region near $r=\infty$ of a KT solution (11) always behaves like an
RNdS geometry with mass equal to the total mass $\sum M_i$. If this
is overmassive, there will be a curvature
singularity in this region, whether the
individual (undermassive) holes have merged or not. This singularity at
the antipodal point of the universe has nothing directly to do with the
black hole merger. To obtain nonsingular initial data we can eliminate
this singularity by replacing it with a collapsing sphere of dust, as in
Sect.\ 4.1. The resulting diagram is shown in Fig.\ 18b. The
initial data induced on the spacelike surface shown by the thick curve
are now nonsingular. In the time development shown, beyond the future
Cauchy horizons $r=0$, a curvature singularity appears. It comes in
from infinity through the infinite throats of the merged black holes
and spreads to $r'=0$, i.e. to the antipodal point of the universe.
The fact that the infinite throats are hidden behind trapped surfaces
does not seem to be sufficient to prevent the singularity from coming
``out of'' the throat. Perhaps a less coordinate-oriented way of saying
this is that all of space collapses down the throats, carrying all
observers with it.

It is clear from the diagram that all observers originating on the initial
surface will reach the Cauchy horizon, and if they extend beyond, they
will see the singularity. We have seen that the Cauchy horizon surrounding
a typical KT throat is not smooth, so that delicate observers may not
survive the crossing. But by distributing several KT masses symmetrically
about a given one, we can make that one as differentiable as necessary
to ensure an observer's survival. So it is reasonable to conclude that
cosmic censorship is violated in these examples.

The initial data in these examples already contain the black holes' infinite
throats, and are not compact. Can we first form the black holes from
collapsing dust, and then let them go through the above scenario? We have
seen in Fig.\ 10 that we can have simultaneous collapse of a dust ball to
form a black hole, and simultaneously remove the overmassive singularity
at the antipodal point by another dust ball. The problem is now that in
the KT solution, as in the RNdS case shown in Fig.\ 10, the two balls
collide before any singularities have formed, at least if the balls move
on $r=0$ resp.\ $r'=0$ trajectories in KT (cosmological) coordinates. Even
if we allow more general trajectories we know for charged test particles that
the trajectories tend to avoid singularities of the same charge; and even
if naked singularities were formed later in the evolution, one would not
know whether they were a fundamental property of the theory, or due to
the dust approximation (``shell crossing singularities,'' which occur
also in the absence of gravity and hence have nothing to do with cosmic
censorship).

Even if we accept the KT solution's infinite throats in place of compact
initial data, we still do not yet have a serious
violation of cosmic censorship, because the general KT solution is still
quite special. The initial position and masses can be specified arbitrarily,
but not the initial velocities. The constraints on the initial values
can be solved in more general (but still not quite generic) contexts,
for example one can drop the $Q_i^2 = M_i^2$ condition \cite{BHKT}.
These initial data can be analytic, but we do not know what happens
beyond the Cauchy horizon. In the general KT solution we have seen that
one has to cross the Cauchy horizon to see the naked singularity. It is
not clear whether more generic solutions have a Cauchy horizon with a
stronger singularity than the KT solution. If so, then cosmic censorship
would be preserved.

\def\refname


{\large\bf References}
\begin{thebibliography}{99}
\bibitem{KT} D. Kastor and J. Traschen, Phys. Rev. D{\bf 47} 5370 (1993)
\bibitem{TN} P. Cru\'sciel and J. Isenberg, Phys. Rev. D{\bf48} 1616 (1993)
\bibitem{Cru} P. Cru\'sciel and D. Singleton, Commun. Math. Phys. {\bf147}
137 (1992)
\bibitem{PSH} K. Peeters, C. Schweigert and J. van Holten, {\em Extended
Geometry of Black Holes}, preprint gr-qc/9407006
\bibitem{WA} M. Walker, J. Math. Phys. {\bf 11} 2280 (1970)
\bibitem{KL} K. Lake, Phys. Rev. D{\bf20} 370 (1979)
\bibitem{CABR} B. Carter in {\em Black Holes} ed.\ C. DeWitt and B.DeWitt
(Gordon \& Breach 1973); D. Brill, Phys. Rev. D{\bf46} 1560 (1992)
\bibitem{BH} D. Brill and S. Hayward, Class. Quantum. Grav. {\bf11} 359 (1994)
\bibitem{MM} F. Mellor and I. Moss, Phys. Rev. D{\bf41} 403 (1990) and
Class. Quantum Grav. {\bf9} L43 (1992); also see
P. Brady and E. Poisson, Class. Quantum Grav. {\bf9} 121 (1992); Brady,
N\'u\~nez and Sinha, Phys. Rev. D{\bf47} 4239 (1993); C. Chambers and I. Moss,
Class. Quantum Grav. {\bf11} 1035 (1994)
\bibitem{K?} K. Lake Phys. Rev. D{\bf19} 421 (1979)
\bibitem{BHKT} Brill, Horowitz, Kastor and Traschen, Phys. Rev. D {\bf49}
840 (1994)
\bibitem{xx} See, for example, V. Moncrief and D. Eardley, Gen. Rel. Grav.
{\bf13} 887 (1981); R. Wald {\em General Relativity}, University of Chicago
Press 1984; P. Joshi, {\em Global Aspects in Gravitation and Cosmology},
Oxford 1993 and the references cited there
\bibitem{GF} R. Gregory and R. Laflamme, Phys. Rev. Lett. {\bf70} 2837 (1993)
\bibitem{HH} J. Horne and G. Horowitz, Phys. Rev. D{\bf48} R5457 (1993)
\bibitem{JAW} J. A. Wheeler {\em Geometrodynamics}, Academic Press 1962
\bibitem{RO} L. Romans  Nucl. Phys. {\bf B383} 395 (1992)
\bibitem{BA} Banks, O'Loughlin \& Strominger, Phys.\ Rev.\ D {\bf 47} 5370
(1993)
\bibitem{ES} G. Ellis and B. Schmidt, Gen. Rel. Grav. {\bf 8}, 915 (1977)
\bibitem{GH} G. Gibbons and S. Hawking, Phys. Rev. D{\bf 15}, 2752 (1977)
\bibitem{HA} G. Hayward, Phys. Rev. D{\bf 47}, 3275  (1993)
\bibitem{BH} D. Brill and G. Hayward, Phys Rev. D{\bf 50}, 4914 (1994)
\bibitem{W} E. Witten, Commun. Math. Phys. {\bf 80} (1981) 381
\bibitem{GS} D. Garfinkle and A. Strominger, Phys. Lett. {\bf B256}, 146
(1991);
Garfinkle, Giddings and Strominger, Phys. Rev. D{\bf 49}, 958 (1994);
H. Dowker et al, Phys. Rev. D{\bf 49}, 2909 and {\bf 50}, 2662 (1994)
\bibitem{MR} R. Mann and S. Ross, {\em Cosmological production of charged
black hole pairs}, DMPTP/R-95/9, gr-qc/9504015.
\bibitem{DB} D. Brill, Phys. Rev. D{\bf 46}, 1560 (1992)

\end{thebibliography}
\end{document}